\documentclass[final,3p,times,twocolumn]{elsarticle}

\usepackage{amsmath,euscript,psfrag,latexsym}
\usepackage{amsthm}
\usepackage{amssymb}
\usepackage{amsfonts}
\usepackage{bbm,color,amstext,wasysym,subfig,cuted,mathtools}
\usepackage{nomencl}
\usepackage{graphicx}
\usepackage{textcomp}
\usepackage{xcolor}
\usepackage{float}
\def\BibTeX{{\rm B\kern-.05em{\sc i\kern-.025em b}\kern-.08em
		T\kern-.1667em\lower.7ex\hbox{E}\kern-.125emX}}   
\newcommand{\mC}{{\mathbb C}}

\newcommand{\mR}{{\mathbb R}}

\newcommand{\bK}{{\mathbf K}}

\newcommand{\bG}{{\mathbf G}}
\newcommand{\bU}{{\mathbf U}}

\newcommand{\bPsi}{{\boldsymbol \Psi}}
\newcommand{\bC}{{\boldsymbol{\mathcal C}}}

\newcommand{\bX}{{\mathbf X}}
\newcommand{\bY}{{\mathbf Y}}

\newcommand{\bA}{{\mathbf A}}

\newcommand{\cF}{{\mathcal F}}

\makenomenclature

\usepackage{etoolbox}
\renewcommand\nomgroup[1]{%
	\item[\bfseries
	\ifstrequal{#1}{P}{Parameters}{%
		\ifstrequal{#1}{V}{Variables and Functions}{%
			\ifstrequal{#1}{N}{Notations}{}}}%
	]}

\newtheorem{thm}{Theorem}[section]
\newtheorem{assumption}{Assumption}

\journal{IJEPES}

\begin{document}

\begin{frontmatter}



\title{Data-Driven Distributed Voltage Control for Microgrids: A Koopman-based Approach}


\author[inst1,inst2]{Vladimir Toro}
\author[inst1]{Duvan Tellez-Castro}
\author[inst1]{Eduardo Mojica-Nava}
\author[inst2]{Naly Rakoto-Ravalontsalama}




\begin{abstract}
This paper presents a distributed data-driven control to regulate the voltage in an alternate current microgrid (MG). Following the hierarchical control frame for MGs, a secondary control for voltage is designed with a data-driven strategy using the Koopman operator. The Koopman operator approach represents the nonlinear behavior of voltage as a linear problem in the space of observables or lifted space. The representation in the lifted space is used together with linear consensus to design a model predictive control (MPC). The complete algorithm is proved in an MG model including changes in load, transmission lines, and the communication graph. The data-driven model regulates voltage using a distributed approach based only on local measurements, and includes reactive power constraints and control cost minimization.
\end{abstract}




\end{frontmatter}


\mbox{}

\nomenclature[N]{$f$}{Denotes a function representing
	the dynamic of the inverter}
\nomenclature[N]{$\lVert \cdot  \rVert$}{Denotes the Euclidean norm}
\nomenclature[N]{$\mid \cdot \mid$}{Denotes magnitude of a complex number}
\nomenclature[N]{$\dagger$}{Denotes the pseudo-inverse of a matrix}
\nomenclature[N]{$\mathbb{R}$, $\mathbb{C}$}{Denotes the set of real and complex numbers, respectively}
\nomenclature[N]{$\mathbb{R}^{n}$, $\mathbb{R}^{n \times n}$}{Denotes the set of real vectors and matrices, respectively}
\nomenclature[P]{$\mathcal{N}^{i}$}{Denotes the set of neighbors of the $i^{th}$ agent}
\nomenclature[P]{$H_{p}$}{Denotes the prediction horizon for the predictive control}
\nomenclature[P]{$Q^{ref,i}$}{Denotes the reactive power reference at the $i^{th}$ agent}
\nomenclature[P]{$\tau^{i}$}{Denotes the time constant for the low-pass filter at the $i^{th}$ agent}
\nomenclature[P]{$nq^{i}$}{Denotes the droop gain for the reactive power at the $i^{th}$ agent}
\nomenclature[P]{$\mathcal{L}$}{Denotes the Laplacian matrix}
\nomenclature[P]{$Q$}{Denotes the weight for the state at the MPC controller}
\nomenclature[P]{$R$}{Denotes the weight for the input at the MPC controller}
\nomenclature[P]{$S$}{Denotes the weight for the consensus term at the MPC controller}
\nomenclature[P]{$V^{ref}$}{Denotes the voltage of reference}
\nomenclature[P]{$N$}{Denotes the number of the inverters in the microgrid}
\nomenclature[P]{$\mathcal{A}$}{Denotes the lifted state matrix}
\nomenclature[P]{$\mathcal{B}$}{Denotes the lifted input matrix}
\nomenclature[P]{$\mathcal{C}$}{Denotes the projection state matrix}
\nomenclature[P]{$\Psi$}{Denotes the vector of observables}
\nomenclature[V]{$\delta^{i}$}{Denotes the phase angle at the $i^{th}$ agent}
\nomenclature[V]{$V^{ij}$}{Denotes the voltage between agents $i$ and $j$}
\nomenclature[V]{$Y^{ij}$}{Denotes the $ij^{th}$ element of the admittance matrix $Y$}
\nomenclature[V]{$\theta^{ij}$}{Denotes the impedance angle of the load}
\nomenclature[V]{$B^{ij}$}{Denotes the susceptance value between inverters $i$ and $j$}
\nomenclature[V]{$P^{i}$}{Denotes the instantaneous active power at the $i^{th}$ agent}
\nomenclature[V]{$Q^{i}$}{Denotes the instantaneous reactive power at the $i^{th}$ agent}
\nomenclature[V]{$P^{m,i}$}{Denotes the medium active power at the $i^{th}$ agent}
\nomenclature[V]{$Q^{m,i}$}{Denotes the medium reactive power at the $i^{th}$ agent}
\nomenclature[V]{$Q^{L,i}$}{Denotes the reactive power demanded by the load at the $i^{th}$ agent}
\nomenclature[V]{$u^{i}$}{Denotes the secondary control input at the $i^{th}$ agent}
\nomenclature[V]{$x(k)$}{Denotes the state vector at time $k$}

\printnomenclature

\section{Introduction}
\label{sec:Introduction}
The MG is the key concept for the organized integration of renewable and distributed energy resources. It is based on inverters that work in grid-forming configuration when the MG is isolated from the utility network \cite{Rocabert2012}. Hence, voltage and frequency references are generated and maintained by the MG itself \cite{Sen2018}. In particular, voltage regulation is a challenging problem due to its variations along transmission lines, the effects of nonlinear loads such as rectifiers, and the effect of reactive power regulation. The proper control in a MG relies in the correct modeling of the inverters and the effect of the transmission lines. In \cite{Pogaku2007}, a linear small-signal model of the whole MG is presented, including the transmission line effect, while in \cite{Dorfler2016} an individual approach  for modeling inverters is presented. A complementary approach is presented in \cite{Schiffer2016}, which extends the particular model to a generic topology and an arbitrary number of inverters. In all cases, several generalizations and approximations should be done due to the inherent differences in time-constants, tolerances, gains, and impedance of each inverter. Therefore, the controller designing and optimization of a MG lacks of some parameters and behaviors for a more detailed model. This gap can be fulfilled by using data measured from each inverter, which allows modelling the system including several parameters such as the effect of different tolerances at each inverter.

Voltage regulation in islanded MGs depends on the impedance among inverters and the reactive power balance. The type of the power sources and the capacity of the storage devices determine the availability of active and reactive power. Following the hierarchical control architecture for MGs\cite{Guerrero2011}, \cite{Guerrero2013}, the secondary control must keep the voltage at the reference value while keeping the power-sharing condition. The last restriction is usually maintained by primary control using a decentralized droop-control strategy. However, it introduces a trade-off between voltage regulation and power-sharing; i.e., an improvement in voltage regulation degrades power-sharing and, in contrast, a better power-sharing deteriorates voltage regulation. A proper inverter control might require solving an optimization problem where the objective function has to minimize the difference between the voltage measured at the inverter's output and the reference value while reducing the control effort. In particular, MPC has been successfully applied to several control problems \cite{Agachi}. Voltage regulation problem and optimization in MGs is presented in works such as \cite{Anderson2019}, \cite{Qu2020}. In \cite{Anderson2019}, a nonlinear relation voltage-reactive power is used to design a distributed secondary control. However, solving the non-convex optimization problem at each sampling time requires specialized solvers, which might increase computational time. In \cite{Guo2019}, it is proposed a distributed voltage controller that includes a distributed information system (DIS) and a distributed MPC (DMPC). The DIS synchronizes the voltages, while the DMPC corrects the voltage deviations while keeping the power sharing; this model uses a linearized model for the voltage predictions. In \cite{Ge2021},  an event-triggered DMPC voltage controller for a MG  is proposed.  It is based on a linear model of the inverter and a nonasymptotic inverter to manage the nonlinear behaviors; the DMPC includes a consensus part in the cost function. This approach has the advantage of reducing the computational cost while improving the performance of the communication system; however, it relies on feedback linearization of the inverters. In \cite{RangaSaiSesha2018}, a centralized MPC improved by using a Laguerre polynomials is proposed; the MPC is based on a state space model of the MG that includes the dynamics of the components and the transmission lines between sources. In contrast, our approach uses a Koopman-based linear predictor from the equation that includes droop control to keep the power sharing condition, and a noncooperative distributed control is used in the MPC problem.

The reduction in the cost of sensors and the growth in the capability of collecting measurements from a system have made available a huge quantity of data. In recent years, several techniques have been developed for data-analysis and control designing based on least-squares regression to approximate the Koopman operator \cite{Rowley2014}. The Koopman operator allows representing a nonlinear system as a linear system of infinite dimensions in the space of observables, which is a new space defined by a set of functions or bases in the also called lifted space. The Koopman representation of a system is determined by the Koopman mode decomposition (KMD) \cite{Susuki2015}, that defines a set of associated eigenfunctions, eigenvectors, and modes. This representation can be found analytically in particular cases; however, the nonlinear nature of a system can make it intractable. The KMD of a system can be approximated by data-based techniques using only measurements from the system. Two of the well-known and established data-based algorithms to find the KMD of a system are the dynamic mode decomposition (DMD) \cite{Proctor2016}, and the more general extended dynamic mode decomposition (EDMD) \cite{williams2015data}. This linear representation is very suitable for MPC designs where the restriction given by the voltage-reactive power relation can be represented linearly in the lifted space.

Koopman operator has been used for power stabilization in systems with synchronous generators. In \cite{Korda2018}, a linear predictor is designed using Koopman operator to regulate the frequency of the system. In \cite{Ping2021}, a group of synchronous generators is represented in the Koopman space where the set of basis functions is determined by a deep neural network; thus transient stability is improved using model predictive control. On the other hand, controllers designed using the Koopman approach are not set in a distributed form. It is worth noticing that the coupled nature of power systems implies that changes in an inverter generate variations over the whole system. A decentralized controller corrects its voltage only based on its own measurements; however, this solution might not be optimal. As a response to this problem, the optimization might be centralized, but it implies several issues due to the cost of a full communication system \cite{Yang2019}. Besides, a distributed approach only needs local information defined by a set of neighbors at each agent; it also makes the system robust to failures in the communication system. Thus, the linear representation based on the Koopman operator simplifies the optimization problem and simplifies the design of a distributed model predictive control (DMPC).

The main contributions of this work are two-fold:\\
1) A non-cooperative distributed controller to regulate the voltage in a MG is designed. Using the EDMD algorithm, the Koopman operator is approximated and used to design a linear observer. The objective function is quadratic with a restriction that includes an additional term for the noncooperative distributed term.

2) The convergence of the distributed non-cooperative algorithm is proved, and it is compared with a nonlinear MPC showing that the proposed Koopman-based approach is faster. 

The rest of the paper is organized as follows: in Section 2, the concepts for the MG model, the MG control, the MPC algorithm, and an introduction to the Koopman operator are presented. Section 3 presents the distributed MPC controller for voltage using the Koopman operator, and a stability analysis of the proposed control law. In Section 4, simulation results are shown including the Koopman representation of inverters and some scenarios for the MG. Finally, conclusions and future work are drawn in Section 5.


\section{Preliminaries}
\label{sec:Preliminaries}
This section presents the fundamentals of MG modeling using droop-control and the general concept of the Koopman operator for data-driven representation of dynamical systems using EDMD.

\subsection{Microgrid Modeling}
MGs are modeled by the power flow between inverters through transmission lines. Admittance values are used instead of impedance values for ease of notation. For an impedance between two nodes denoted by $Z^{ij}$, the admittance value is given by $Y^{ij}=G^{ij}+B^{ij}$ where $G^{ij}$ is the conductance, and $B^{ij}$ is the susceptance. Active and reactive power flow between inverters depends directly on voltage magnitude and the phase angle as given by the Kundur equation for $N$ interconnected inverters as follows

\begin{equation}
	\label{powerequationP}
	P^{i}=\sum_{j \in \mathcal{N}^{i}} V^{i}V^{j}|Y^{ij}|\cos(\theta^{ij}+\delta^{j}-\delta^{i}),
\end{equation}
\begin{equation}
	\label{powerequationQ}
	Q^{i}=\sum_{j \in \mathcal{N}^{i}} V^{i}V^{j}|Y^{ij}|\sin(\theta^{ij}+\delta^{j}-\delta^{i}),
\end{equation}
where $V_{i}$, $V_{j}$ are the magnitudes of the voltages at nodes $i$ and $j$, respectively. $|Y_{ij}|$ is the magnitude of the admittance between nodes $ij$, $\mathcal{N}^{i}$ is the set of neighbors for the $i^{th}$ inverter, $\theta_{ij}$ is the admittance angle, $\delta_{i}$, $\delta_{j}$ are the voltage phase angles at nodes $i$, $j$, respectively.

The MG model given in (\ref{powerequationP}) and (\ref{powerequationQ}) is simplified by assuming a very low resistance value in the transmission lines of the form $Y^{ij}=jB^{ij}$ with $j=\sqrt{-1}$, $\theta^{ij}=\frac{\pi}{2}$. Last condition can be achieved by using a virtual impedance feedback control or by simply putting a large impedance value at the inverter's output \cite{Zhong2016}. Assuming a negligible resistance value (\ref{powerequationP}) and (\ref{powerequationQ}) become \cite{Schiffer2016}, \cite{Schiffer2014}
\begin{equation}
	\label{powerequationPsimple}
	P^{i}=\sum_{j \in \mathcal{N}^{i}}|B^{ij}| V^{i}V^{j}\sin(\delta^{j}-\delta^{i}).
\end{equation}
\begin{equation}
	\label{powerequationQsimple}
	Q^{i}=(V^{i})^{2}\sum_{j\in \mathcal{N}^{i}}|B^{ij}|-\sum_{j \in \mathcal{N}^{i}} V^{i}V^{j}|B^{ij}|\cos(\delta^{j}-\delta^{i}).
\end{equation}
Based on (\ref{powerequationPsimple}) and (\ref{powerequationQsimple}) active power depends on the phase changes, while reactive power depends mainly on the voltage changes. This is particularly useful because it allows decoupling the active and reactive power control. However, this approximation does not always hold and might be affected by the type of loads in the system, such as nonlinear ones \cite{Zhong2016}. 
 
It is important to consider the quadratic term $(V^{i})^{2}$ corresponding to the reactive power generated by the own inverter. Reactive power expression (\ref{powerequationQsimple}) is simplified by considering the difference $\delta_{j}-\delta_{i}$ small enough to be zero as follows

\begin{equation}
	\label{powerequationQsimpleCos}
	Q^{i}=\sum_{j\in \mathcal{N}^{i}}(V^{i})^{2}|B^{ij}|-\sum_{j \in \mathcal{N}^{i}} V^{i}V^{j}|B^{ij}|.
\end{equation}

\subsubsection{Primary Control}
For primary control, the most common technique is conventional droop control. This control technique subtracts from the voltage of reference, a value proportional to the reactive power measured $Q^{m}$. Due to the rapid variations on the power measured, it is necessary to work with a medium value, which is generated by a low-pass filter whose time-value $\tau$ varies between tens of milliseconds to hundreds of milliseconds. The droop-equation is given by
\begin{eqnarray}
	\label{droopequation}
	V^{i}=V^{ref}-nq^{i}Q^{m,i},
\end{eqnarray}
where $nq^{i}$ is the droop-coefficient for reactive-power, and $Q^{m,i}$ is the medium reactive-power given by
\begin{eqnarray}
	\label{lowpassfilter}
	Q^{m,i}=\dfrac{1}{1+\tau^{i} s}Q^{i}.
\end{eqnarray}

The relation between instantaneous reactive-power and the medium reactive power is derived from (\ref{lowpassfilter}) to the next differential equation
\begin{equation}
	\label{lowpassdifferential}
	Q^{m,i}+\tau^{i}\dot{Q}^{m,i}=Q^{i}.
\end{equation}
Combining (\ref{powerequationQsimpleCos}), (\ref{droopequation}), and (\ref{lowpassdifferential}) the next differential equation for voltage is generated
\begin{equation} 
	\label{eqn:voltagenonlinear}
	\begin{split}
		\tau^{i}\dot{V}^{i}&=-V^{i}+V^{ref,i}-nq^{i}(
		Q^{L,i}+V^{i}V^{i}\sum_{i\in \mathcal{N}}|B^{i,j}|\\
		& -\sum_{i\in \mathcal{N}}V^{i}V^{j}|B^{i,j}|-Q^{ref,i}),
	\end{split}
\end{equation}
where $Q^{L,i}$ is the power load, and $Q^{ref,i}$ is the power reference.

Droop control deviates voltage and frequency from their reference values. These limitations are overcome by including a secondary control layer that can be centralized or distributed \cite{Sen2018}, \cite{Dorfler2016}. The secondary control keeps the voltage inside the reference values while maintaining the power-sharing condition (each inverter supplies power according to its maximum capacity). However, these two conditions are opposed, which creates a trade-off between them.

In any case, voltage and current controllers and droop-control have several limitations, such as that their performance depends on the control parameters; their sensibility to load changes; the tuning process can be challenging, and the necessity of a pulse-width modulation (PWM) stage. MPC allows overcoming those limitations by performing an optimization process on finite time. In inverter-based MGs, predictive control can be made local for converter-level or globally, also known as grid-level \cite{Hu2021}.

MPC can be classified either as continuous control set (CCS-MPC) or finite control set (FCS-MPC) for control-level in MGs. CCS-MPC generates continuous signals for the PWM converters, while FCS-MPC generates discrete-time signals; thus PWM signals are not necessary \cite{Hu2021}.

The inverter dynamic is represented as follows
\begin{equation}
	\label{eq:convertermodel}
	x(k+1)=f(x(k),u(k)),
\end{equation}
where $x(k)$ represents the set of states for the converter at time $k$, $u$ is the control input, and $f$ is a function representing the dynamics of the inverter.

For a converter modeled with $f$ given by a linear function, system (\ref{eq:convertermodel}) is written as 
\begin{equation}
	\label{eq:converterlinealmodel}
	x(k+1)=Ax(k)+Bu(k),
\end{equation}
where $A\in \mathbb{R}^{n \times n}$ is the matrix of states, $B \in \mathbb{R}^{m}$ is the matrix with the control inputs.

One of the control goals is to reach and keep the state variables or output of the system in their reference values. If reference values are denoted by $x^{*}$ the next condition represents the control goal
\begin{equation}
	\label{eq:referencevalues}
	x(k+1)=x(k)=x^{*}.
\end{equation}
Then, a cost function $L(x,u)$ can be set
\begin{equation}
	\label{eq:costfunction}
	L(x, u)=||x-x^{*}||^{2}_{Q}+||u-u^{*}||^{2}_{R} 
\end{equation}
where $Q \geq 0$ is positive semi-definite and $R>0$ is positive-definite, and $x^{*}$, $u^{*}$ are reference values for the state and input, respectively.

Combining (\ref{eq:converterlinealmodel}) and (\ref{eq:referencevalues}) the steady-state input can be determine as
\begin{equation}
	\label{eq:optimalsolution}
	x^{*}=(I-A)^{-1}Bu. 
\end{equation}

Usually, the solution obtained in the optimization process $u^{opt}$ does not belong to the finite set of inputs for the model predictive problem.

\subsubsection{Model predictive control design}
Secondary control should correct the voltage deviations while maintaining the power-sharing condition. A local optimization process can correct voltage deviations while reducing the control effort; this is presented as an MPC problem with restrictions
\begin{eqnarray} 
	\min_{u^{i}_{t+k}}& \sum_{k=0}^{H_{p}} ||V^{i}_{t+k}-V^{ref,i}_{t+k}||^{2}_{Q}+||u^{i}_{t+k}||^{2}_{R}
\end{eqnarray}
\begin{eqnarray} 
	\textrm{s.t.}& 
	\begin{split}
		\label{discreteformmain}
		V_{k+1}^{i}&=V^{i}_{k}+T(-V^{i}_{k}+V^{ref,i}_{k}-nq_{i}(
		Q^{L,i}_{k}+V^{i}_{k} V^{i}_{k}\\ & \sum_{i\in \mathcal{N}^{i}}|B^{i,j}|
		-\sum_{i\in \mathcal{N}^{i}}V^{i}_{k}V^{j}_{k}|B^{i,j}|-Q^{ref,i}_{k}))
	\end{split}
\end{eqnarray}
\begin{eqnarray}& 0.95V^{ref}\leq V_{k}^{i}\leq 1.05V^{ref}
\end{eqnarray}
where condition (\ref{discreteformmain}) is the discrete form of (\ref{eqn:voltagenonlinear}) with $T$ the sample time used for the discretization, and $H_{p}$ is the prediction horizon for the MPC problem.   

This optimization problem is solved at each sampling time iteration, and the restriction $(\ref{eqn:voltagenonlinear})$ is nonlinear, which implies a non-convex problem with more computational effort and possible convergence issues. These non-linearities can be managed by using a data-based approach. Usually, series of voltage and current measurements from each inverter are available from sensors along the MG. The next section presents the Koopman operator and a data-based approach for nonlinear systems representing.

\subsection{Koopman Operator and EDMD}
The Koopman operator is a linear operator acting on a set of observables generating a linear representation of a nonlinear dynamical system but of infinite dimension. Several data-based algorithms have been developed to get the Koopman representation of dynamical systems, such as Galerkin Projections, Laplace averages, and EDMD with its variants \cite{williams2015data}. The EDMD algorithm is a generalization of DMD suited to get the Koopman representation of nonlinear systems.

Numerical approximation of Koopman operator to systems with inputs are extensions of EDMD method. For control-affine systems, it is possible to find a bilinear approximation in the lifted space or assume the input as a parameter; hence it may get different Koopman representations for each input. It can be assumed that the input has a linear relation with the system in the lifted space, and thus it is possible to use the traditional linear control techniques for nonlinear systems in the lifted space. To build the model in  lifted space, a brief definition of the Koopman operator in discrete time is presented.

Consider a dynamical system of the form 
\begin{eqnarray} 
	x_{k+1} = T(x_k), \nonumber
\end{eqnarray}
where $T: X\to X$ and $X$ is the state space. It is assumed that there is a vector space of observable ${\cal F}$ with $\psi:X\to \mC$ such that for $\psi\in {\cal F}$, $\psi\circ T\in \cal F$, then the Koopman operator ${\cal K}: \cF\to \cF$ can be defined as
\begin{eqnarray}
	[{\cal K} \psi](x)=\psi(T(x)). \nonumber
\end{eqnarray}
The eigenfunction $\varphi_\lambda$ associated with  eigenvalue $\lambda\in \mC$ for the Koopman operator is defined as 
\begin{eqnarray}
	\label{eigenfunction_koopman} \nonumber
	[{\cal K}\varphi_\lambda](x)=e^{\lambda t}\varphi_\lambda(x)
\end{eqnarray}

The goal is to find a finite numerical approximation of ${\cal K}$ denoted as $\bK$, only with data available from computational simulation or physical measures. To this end, consider the snapshots of data as
\begin{eqnarray}
	\bX=[x_0,\ldots,x_{M}],\;\;\;\bY=[y_0,\ldots, y_M]. \nonumber
\end{eqnarray}
It is assumed that  a set of linearly independent basis functions $\psi_i\in\cF$ for $i=1,\ldots,N$ exists, and define
\[\cF_N={\rm span}\{\psi_1,\ldots,\psi_N\}.\]
Using the EDMD algorithm, it is possible to construct the finite-dimensional approximation of Koopman operator $\bK:\cF_N\to \cF_N$ as the solution of the following least square problem
\begin{eqnarray}
	\min_{\mathbf{K}\in \mR^{N\times N}}\parallel \mathbf{G} \mathbf{K}- \mathbf{A} \parallel_F\label{compute_K}\nonumber
\end{eqnarray}
where 
\[\bG=\frac{1}{M}\sum_{k=0}^M \bPsi(x_k)\bPsi^\top(x_k),\;\;\bA=\frac{1}{M}\sum_{k=0}^M \bPsi(x_k)\bPsi^\top(y_k)\]
and 
$\bPsi(x)=[\psi_{1}(x),\ldots, \psi_{N}(x)]^\top$ are the basis functions, and $F$ stands for the Frobenius norm. The least-square problem can be solved explicitly with a pseudo-inverse operation as follows
\begin{eqnarray} \label{eq:K}
	\mathbf{K}=\bG^\dagger \bA \nonumber
\end{eqnarray}
where $\dagger$ denotes the pseudo-inverse.

For a system with control inputs, and dynamics as defined in (\ref{eq:convertermodel}). The Koopman operator for the system with control inputs can be defined as
\[ [{\cal K}_c \psi](x_k,u_k) = \psi(T(x_k,u_k),u_{k+1})\]
assuming a data vector of inputs $\bU =[u_0,\ldots,u_{M}]$. Therefore, the EDMD algorithm using the linear basis (i.e., DMD) with inputs can be obtained solving the following optimization problem
\begin{align*}
	\min_{A,B}  \sum_{i=0}^M \parallel y_i - Ax_i -Bu_i \parallel^2.
\end{align*}
Similarly, if it is considered that the control input acts in a linear form, then it can be obtained a linear predictor by solving the following optimization problem
\begin{eqnarray}
\label{AB_identify}\nonumber
\min_{\mathcal{A},\mathcal{B}} \rVert \bPsi(\bY) -\mathcal{A} \bPsi(\bX) - \mathcal{B} \bU \rVert_F.
\end{eqnarray}
where $\mathcal{A} \in \mathbb{R}^{N \times N}$, $\mathcal{B} \in \mathbb{R}^{N \times m}$, for more details see \cite{proctor2018generalizing}, \cite{korda2018linear}. Note that the ${\cal A}$ identified using the above least square problem will correspond to the transpose of $\bK$. To come back to state space is necessary to solve 
\[\min_\bC \rVert \bX - \bC \bPsi(\bX) \rVert_F\]
The matrix $\bC$ also can be computed explicitly as
\[ \bC = \bX \bPsi(\bX)^\dagger, \]
where $\bC \in \mathbb{R}^{n \times N}$.

The next section presents the design of the MPC secondary control, and its distributed approach using the Koopman-based representation.

\section{Problem statement}
This section presents a MPC controller for a MG that uses a Koopman-based model as a predictor. The complete MG model consists of $N$ inverters represented by the nonlinear equation (\ref{voltagenonlinear}), that it is used to generate the necessary data to determine the Koopman representation of each inverter.

\subsection{Data-Driven Secondary Control}
The dynamical model (\ref{voltagenonlinear}) represents the relation between voltage and reactive power, including the admittance between inverters. A secondary control term $u_{i}$ is included as follows \cite{Anderson2019}
\begin{equation} 
\label{voltagenonlinear}
\begin{split}
\tau_{i}\dot{V}^{i}&=-V^{i}+V^{ref,i}+u^{i}-nq_{i}(
Q^{L,i}+V^{i}V^{i}\sum_{i\in N}|B^{i,j}|\\
& -\sum_{i\in N}V^{i}V^{j}|B^{i,j}|-Q^{ref,i}),
\end{split}
\end{equation}
where the terms $V^{i}V^{i}$ and $V^{i}V^{j}$ introduce nonlinearities, that makes the optimization problem non-convex. It also requires extra time processing and a different solver than the linear-based ones.  

The Koopman operator allows representing (\ref{voltagenonlinear}) as a linear system in the space of observables by using the EDMD algorithm. System (\ref{voltagenonlinear}) is simulated and voltage measurements from each inverter are gathered. The load changes are considered perturbations inside the maximum load limit. Different trajectories of the system are generated by changing the initial conditions inside the range $[0,v^{ref}]$.

\begin{figure}[htp]
	\centering
	\includegraphics[width=0.99\linewidth]{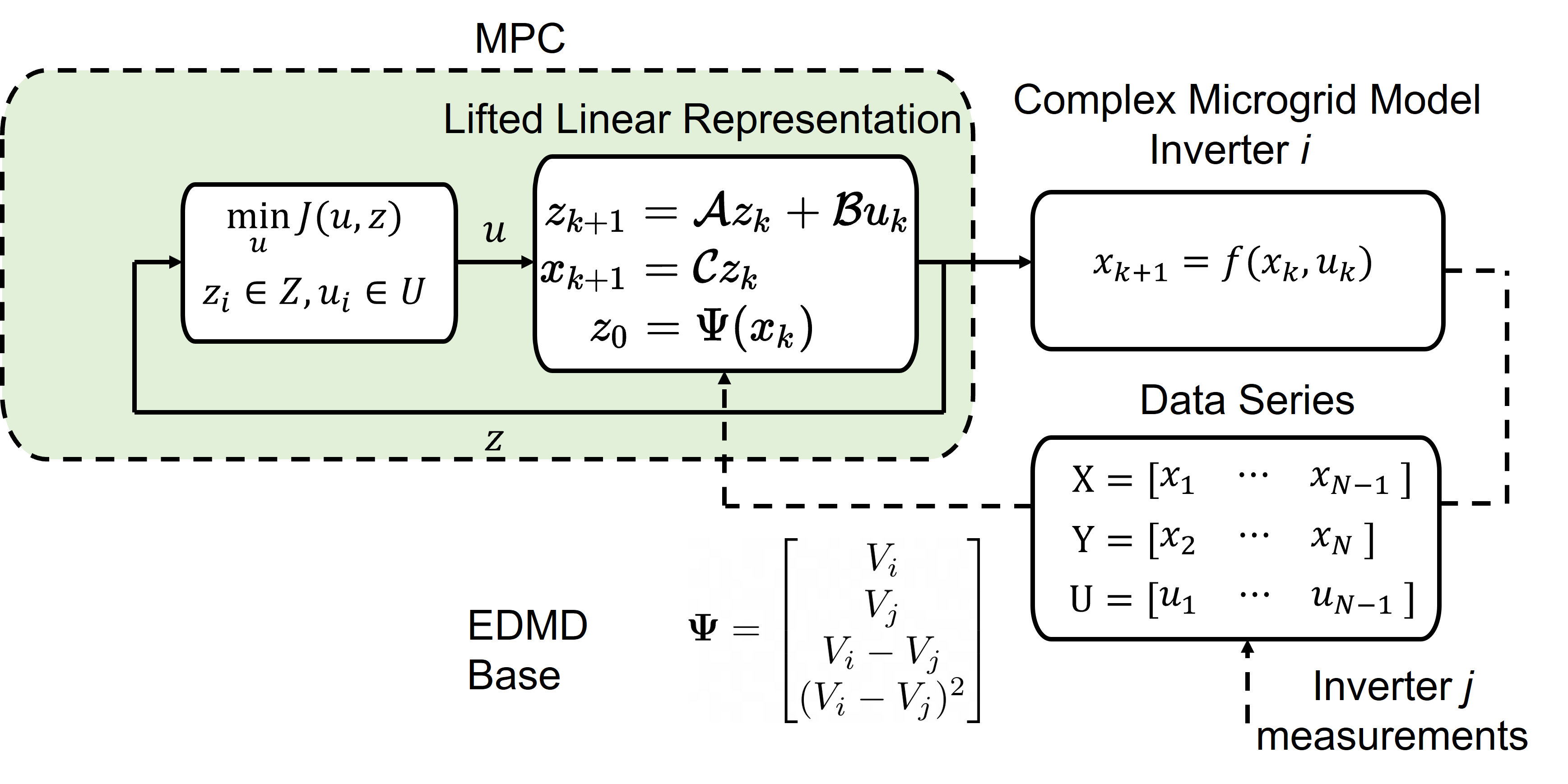}
	\caption{Decentralized Koopman MPC Scheme.}
	\label{Fig:KoopmanMPCSingle}
\end{figure}

Figure~\ref{Fig:KoopmanMPCSingle} shows the general schematic design of the Koopman-based MPC controller,  which uses four observables or bases for the EDMD offline algorithm to determine matrices $\mathcal{A}$, $\mathcal{B}$, and $\mathcal{C}$. The MPC works in a decentralized mode without sharing information with other inverters, and just using the voltage measurements of neighbor inverters for the Koopman model. 

\subsection{Distributed Koopman MPC}
The natural coupling of power systems implies that any change over the components of the MG changes the voltage and phase values of the whole system. Thus, a complete decentralized controller is not practical to find the optimal values to reach the voltage and frequency reference values. In a networked system, the optimization process can be done by a centralized or distributed controller. The last one is more practical due to its flexibility in the communication system design and its reliability to failures. 

Following a distributed approach, the MPC design based on the linear representation (\ref{voltagenonlinear}) in the lifted space is complemented by the consensus agreement among the voltage measurements coming from the set of neighbors of the $i^{th}$ agent. Then, the predictive controller also reduces the differences among the voltages of the inverter as follows
\begin{eqnarray} 
	\min_{u^{i}_{t+k}}& \sum_{k=0}^{H_{p}} |V^{i}_{t+k}-V^{ref,i}_{t+k}||^{2}_{Q}+||u^{i}_{t+k}||^{2}_{R}
	\label{Eq:linearmpc1}
\end{eqnarray}
\begin{eqnarray} 
	\textrm{s.t.}& 
	V^{i}_{k+1}=V^{i}_{k}+T(\mathcal{A}V^{i}_{k}+\mathcal{B}u^{i}_{k})+S\mathcal{L}(i,:)\mathbf{V}_{i}
	\label{Eq:linearmpc2}
\end{eqnarray}
\begin{eqnarray}& 0.95V^{ref}\leq V_{k}^{i}\leq 1.05V^{ref}
	\label{Eq:linearmpc3}
\end{eqnarray}
where $\mathcal{L}(i,:)$ corresponds to the $i^{th}$ row of the Laplacian matrix, $\mathbf{V}_{i}=[V_{1} \quad V_{2} \quad \ldots \quad V_{i} \quad \ldots \quad V_{n}]^{\top}$ is a column vector with the voltage values of each agent, and $S$ is the gain value for the consensus term.

\begin{figure}[htp]
	\centering
	\includegraphics[width=1.0\linewidth]{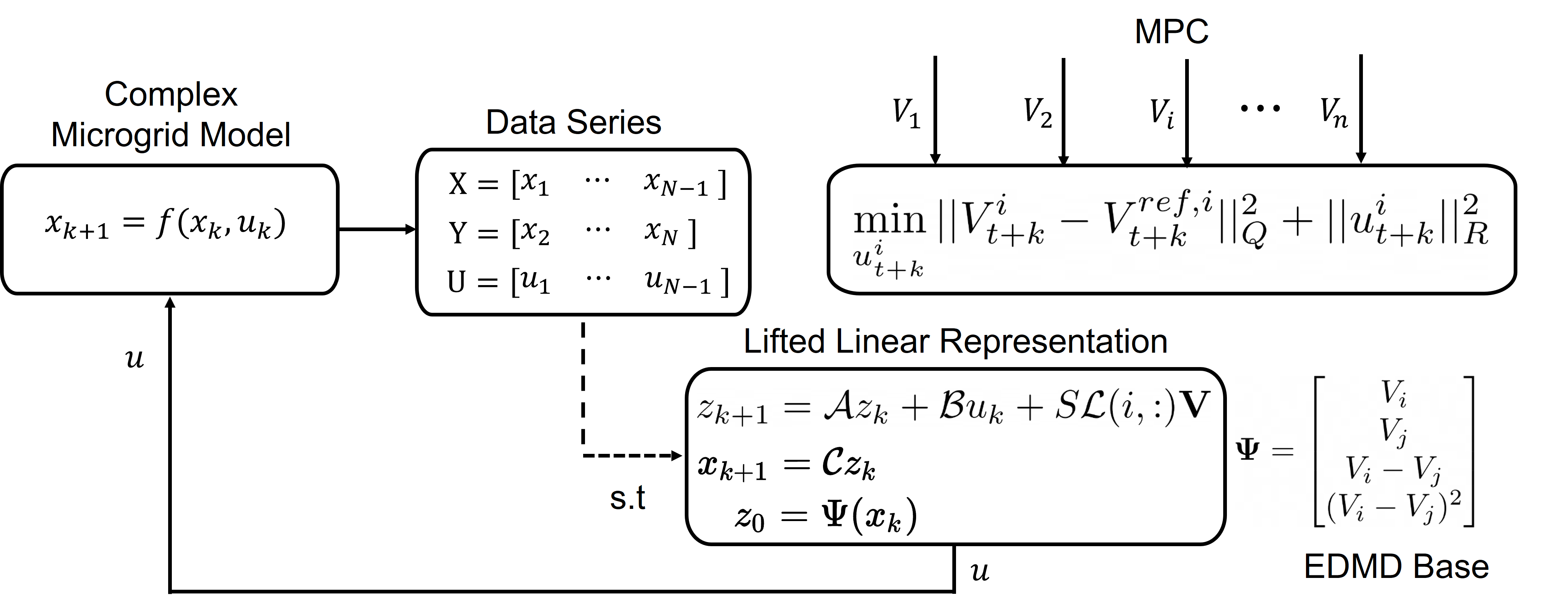}
	\caption{Distributed Koopman MPC Scheme.}
	\label{Fig:distributedKoopmanMPC}
\end{figure}

Figure~\ref{Fig:distributedKoopmanMPC} shows the general scheme for the DMPC. Voltage measurements are gathered for the set of neighbors of the $i^{th}$ agent. Then the four based functions are used to determine the Koopman operator. The MPC is designed to minimize the cost function with the restriction given by the linear representation obtained by the Koopman operator and the consensus term given by the Laplacian matrix. 

Some aspects of the convergence of the algorithm are considered in the next subsection. The linear representation in the lifted space allows using already established conditions for MPC design and convergence.

\subsection{Convergence Analysis}
The convergence of the control algorithm defined by (\ref{Eq:linearmpc1})-(\ref{Eq:linearmpc3}) is determined by the behavior of each inverter represented by the linear matrices $\mathcal{A}$, $\mathcal{B}$ in the observables space, and the selection of the MPC parameters $Q$, $R$, and $S$. 

\begin{assumption}
	The graph $\mathcal{G}$ is connected with $\mathcal{L}$ semi positive definite with an eigenvalue $\lambda_{1}=0$.
\end{assumption}

\begin{thm} Optimization problem given by (\ref{Eq:linearmpc1})-(\ref{Eq:linearmpc3}) defined by $Q$, $R$, $S$, and $\mathcal{L}$, with the linear system represented by matrices $\mathcal{A}$, $\mathcal{B}$, and $\mathcal{C}$ is asymptotically stable and converges to the optimal solution.
\end{thm}

\begin{proof}
The cost function and the linear system are defined by
$$L(x,u)=\dfrac{1}{2}x^{\top}Qx+\dfrac{1}{2}u^{\top}Ru \qquad f(x,u)=Ax+Bu+\mathcal{L}\mathbf{x}$$

The optimization problem for $LQ$ regulator is solved by using the calculus of variation. As shown in \cite{Control2012}, the next set of identities should be fulfilled
$$L_{x}=Qx \qquad L_{u}=u^{\top}R \qquad f_{x}=A+\mathcal{L} \qquad f_{u}=B$$
$$\dot{\lambda}^{\top}=-L_{x}-\lambda^{\top}f_{x} \qquad
L_{u}+\lambda^{\top}f_{u}=0$$
Assuming $\lambda=Px$
$$u^{\top}R+\lambda^{\top}B=0 \qquad u=-B^{\top}PxR^{-1}$$
$$P\dot{x}=-Qx-A^{\top}Px-\mathcal{L}Px$$
$$P(Ax+Bu+\mathcal{L}\mathbf{x})=-Qx-A^{\top}Px-\mathcal{L}Px$$
Sustituting $u$
$$PAx+A^{\top}Px+P\mathcal{L}\mathbf{x}+\mathcal{L}Px-PBB^{\top}PR^{-1}x+Qx=0$$
Then, the next Riccati equation is obtained
$$A^{\top}P+PA+Q-PBR^{-1}B^{\top}P+\mathcal{L}P+P\mathcal{L}=0$$

For a control input of the form $u=-Kx$, where $K$ is a square gain matrix given by $K=-ux^{-1}$, and replacing $u=-B^{\top}PxR^{-1}$ we obtain $K=-B^{\top}PR^{-1}$.

For a Lyapunov function of the form $V(x)=x^{\top}Px$, the first derivative of $x$ is given by
$$\dot{V}=\dot{x}^{\top}Px+x^{\top}P\dot{x}$$
$$\dot{V}=[(A-BK)x+\mathcal{L}\hat{x}]^{\top}Px+x^{\top}P[(A-BK)x+\mathcal{L}\hat{x}]$$
Considering that the terms $x$ tends to the reference value, the product $\mathcal{L}\hat{x}\approx \mathbf{0}$, and replacing $K$
$$\dot{V}=x[A-PBR^{-1}B^{\top}P]^{\top}Px+x^{\top}P[A-PBR^{-1}B^{\top}P]x$$
Adding the terms $Q$, and $-Q$ and comparing with the Riccati equation
\begin{equation*}
\dot{V}=x[AP+PA-PBR^{-1}B^{\top}P-Q+Q-BB^{\top}PR^{-1}P]x
\end{equation*}
\begin{equation*}
\dot{V}=x[-Q-PBR^{-1}B^{\top}P]x
\end{equation*}
as matrices $Q\geq 0$, $P>0$, and $R^{-1}>0$ the system fulfills the condition of $\dfrac{d}{dx}V(x)<0$, then the closed system is asymptotically stable. 
\end{proof}

The next section presents some simulation results for the proposed algorithm together with the data-driven simulation to get the Koopman representation of each inverter.

\section{Simulation}
\begin{figure}[htp]
	\centering
	\includegraphics[width=1.0\linewidth]{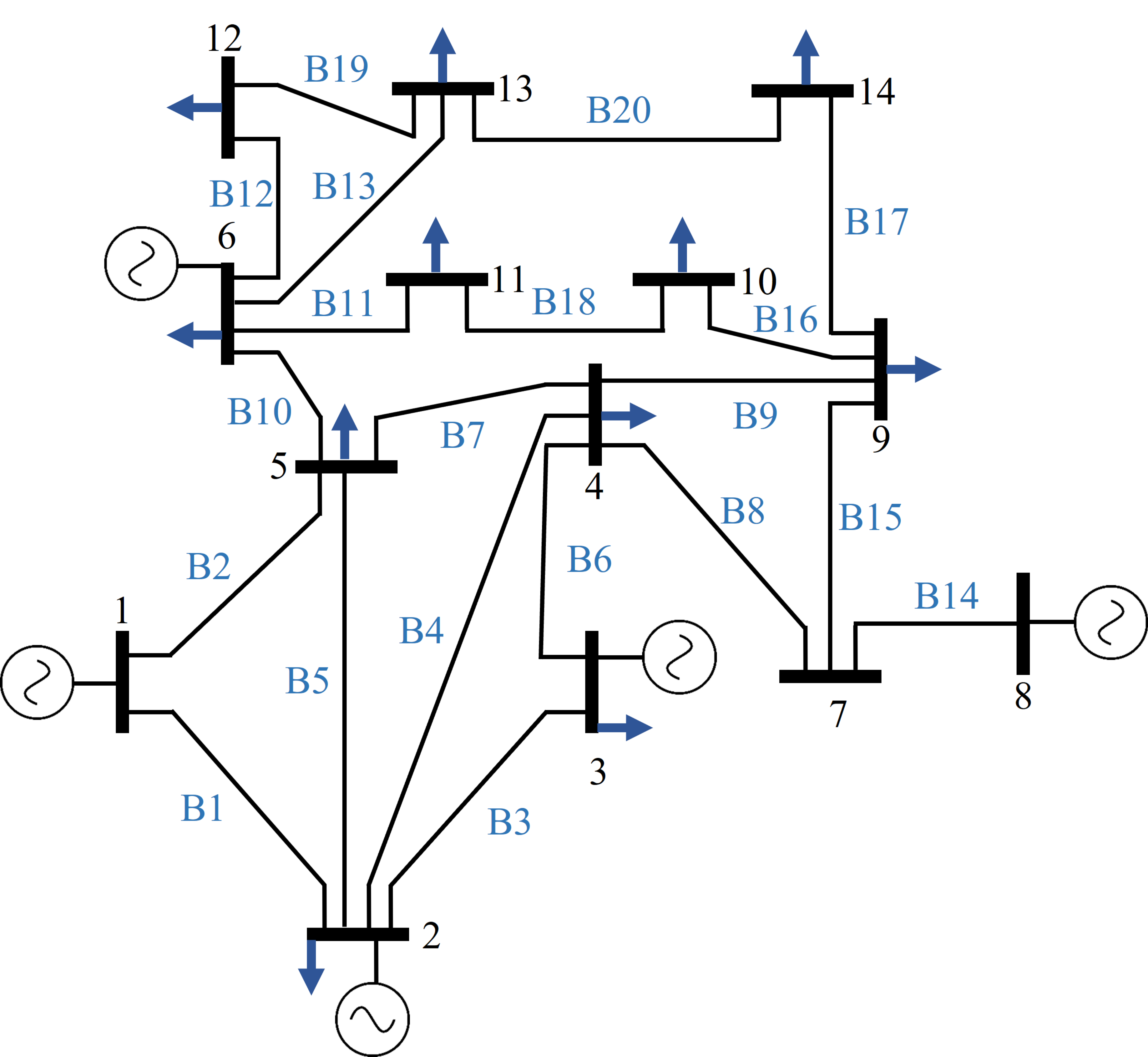}
	\caption{14-nodes IEEE model with five inverters and 11 loads}
	\label{Fig:14nodes}
\end{figure}

In this section, a MG is simulated in Simulink based on the 14-nodes IEEE model \cite{Leon2020}. Figure~\ref{Fig:14nodes} shows the schematic model, which consists of five VSC inverter-based generators, 11 PQ loads, and 20 branches. Transmission lines are denoted from B1 to B20 \cite{Boudreaux2018}. The MPC was simulated in Matlab using Yalmip \cite{yalmip} as the interface and Gurobi as the optimization solver, on an Intel i7-5500U processor at 2.4 GHz and 4GB RAM.

\begin{equation*}
	\mathcal{L}_1=\left[ 
	\begin{array}{ccccc}
		2  & -1   & 0    & -1   &  0 \\
		-1  & 2    & -1   &  0   &  0 \\
		0  & -1   & 3    & -1   & -1 \\
		-1  &  0   & -1   & 2    &  0 \\
		0  & 0    & -1   & 0    &  1
	\end{array}
	\right]
\end{equation*}

Communication among converters is represented by a graph as shown in the left part of Figure~\ref{Fig:GraphChanging} with Laplacian matrix $\mathcal{L}_1$. The graph configuration changes to the one that appears to the right with Laplacian matrix $\mathcal{L}_2$.

\begin{table}[htp]
	\caption{Microgrid Parameters.}
	\centering
	\begin{tabular}{cccccc}
		\hline
		Inverter & $P^{ref}$  & $Q^{ref}$ & $m_{i}$ & $n_{i}$ & $V^{ref}$  \\
	& kW & kvar & $1\times 10^{-4}$ & $1\times 10^{-4}$& rms \\
		\hline
		1 \text{to} 5   &  10  & 5 & 1 & 1 & 120 \\
		\hline
		\label{table:MGparameters}
	\end{tabular}
\end{table}

\begin{table}[htp]
	\centering
	\caption{Transmission line values.}
	\begin{tabular}{clcl}
		\hline
		\begin{tabular}[c]{@{}c@{}}Line \\ 
		\end{tabular} & \multicolumn{1}{c}{\begin{tabular}[c]{@{}c@{}}Inductance\\ (mH)\end{tabular}} & \begin{tabular}[c]{@{}c@{}}Line\\\end{tabular} & \multicolumn{1}{c}{\begin{tabular}[c]{@{}c@{}}Inductance\\ (mH)\end{tabular}} \\ \hline
		B1 & 0.83 & B11 & 2.79 \\ 
		B2 & 3.13 & B12 & 3.59 \\
		B3 & 2.78 & B13 & 1.82 \\ 
		B4 & 2.47 & B14 & 2.47 \\ 
		B5 & 2.44 & B15 & 1.54 \\ 
		B6 & 2.40 & B16 & 1.18 \\ 
		B7 & 0.59 & B17 & 3.79 \\ 
		B8 & 2.90 & B18 & 2.69 \\ 
		B9 & 7.80 & B19 & 2.80 \\ 
		B10 & 3.54 & B20 & 4.88 \\ \hline
	\end{tabular}
	\label{table:inductances}
\end{table}

\begin{table}[htp]
	\caption{Microgrid loads.}
	\centering
	\begin{tabular}{lllllll}
		\hline
		Node                 & 2  & 3  & 4  & 5 & 6 & 9\\
		Load (kW)            & 1  & 1  & 1  & 1 & 1 & 1\\
		\hline
		Node                 & 10 & 11 & 12 & 13& 14&\\
		Load (kW)            & 1   & 1   & 1   & 1  & 1 & \\
		\hline
		\label{table:MGloads}
	\end{tabular}
\end{table}

\begin{table}[htp]
	\centering
	\caption{MPC Parameters.}
	\label{table:MPCparameters}
	\begin{tabular}{llllll}
		\hline
		Paremeter                       & \multicolumn{5}{l}{Inverter} \\
		& 1    & 2    & 3   & 4  & 5    \\ \hline
		State difference gain $Q$       &  1   & 1    &  1  &  1 & 1    \\
		Input gain $R$                  &  5   & 5    & 5   & 5  & 5    \\
		Consensus gain $S$              &  0.2 & 0.2  & 0.2 & 0.2 &0.2  \\ 
		Sampling Time (s)               &  0.1 & 0.1  & 0.1 & 0.1 &0.1  \\
		Voltage restriction $V_{i}(V)$  & \multicolumn{5}{c}{$165\leq V_{i} \leq 175$}    \\
		Control Horizon $H_{p}$         & 3    & 3    & 3   & 3   & 3   \\ \hline
	\end{tabular}
\end{table}

The parameters of the inverters of the MG are summarized in Table~\ref{table:MGparameters}. Active and reactive power values are identical for each inverter together with the droop coefficients with the idea of visualizing the performance of the controller. In Table~\ref{table:inductances} appears the values of inductance for the transmission lines. The values of the loads for active and reactive power are shown in Table~\ref{table:MGloads}. Finally, the general MPC parameters for each local controller are shown in Table~\ref{table:MPCparameters}. The sample time for the controller corresponds with the filter constant value $\tau$, and the control horizon is limited to a few steps ahead.

\subsection{Koopman Algorithm Simulations}
The voltage is measured at the ouput of each inverter and it is sampled at $T_{s}=1$ms for a 10-second-window; the input varies each 1.7s randomly from -1V to 1V. 

\begin{figure}[htpb]
	\centering
	\includegraphics[width=1.0\linewidth]{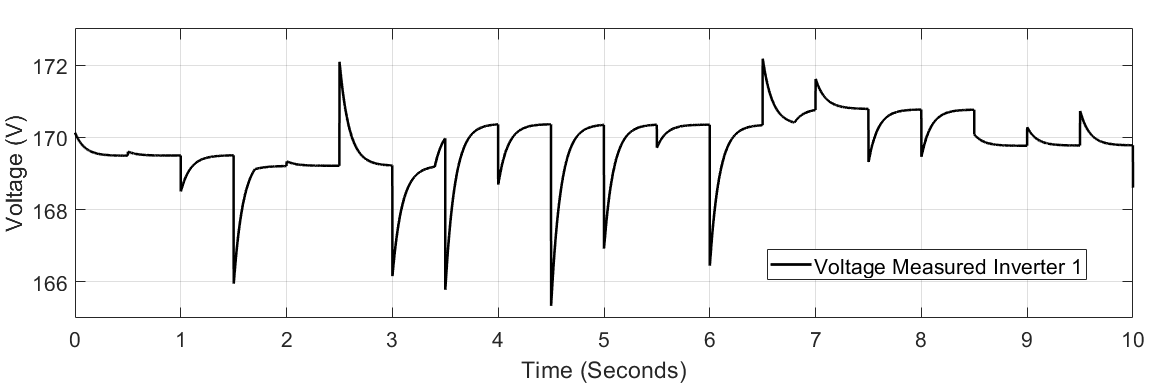}
	\includegraphics[width=1.0\linewidth]{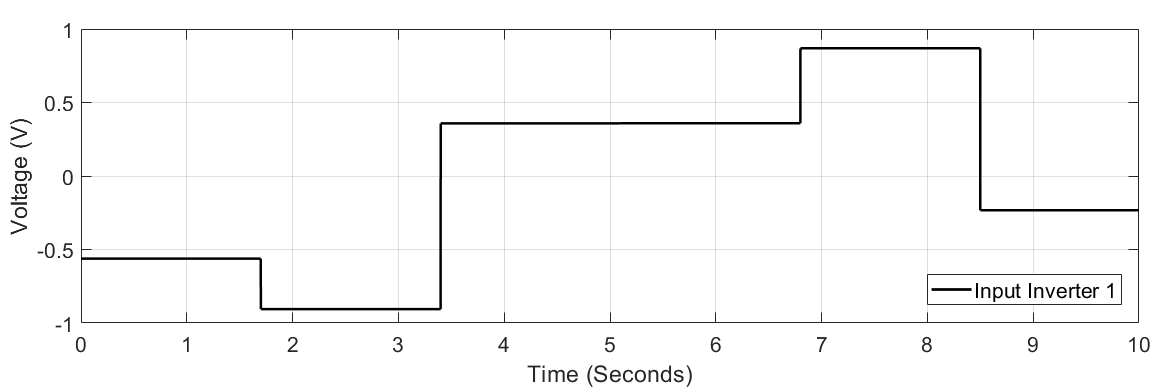}
	\caption{Data Generated to obtain the Koopman representation of inverter 1.}
	\label{Fig:DataInverterOne}
\end{figure}

The data generated to obtain the Koopman representation of inverter one with several different initial conditions and changing the input, are shown in Figure~\ref{Fig:DataInverterOne}. The data gathered is split in two sets, one to obtain the Koopman matrices $\mathcal{A}$, $\mathcal{B}$, and $\mathcal{C}$ and one to verify the validity of the approximation. 

\begin{figure}[htpb]
	\centering
	\includegraphics[width=1.0\linewidth]{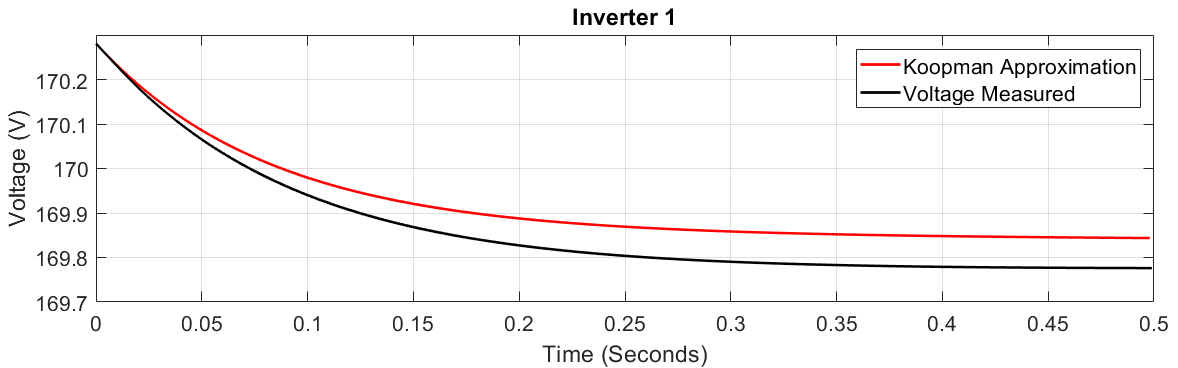}
	\includegraphics[width=1.0\linewidth]{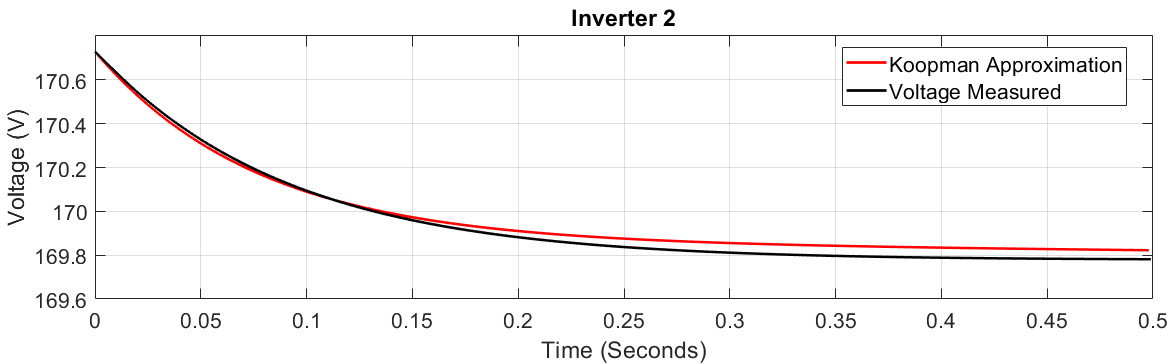}
	\includegraphics[width=1.0\linewidth]{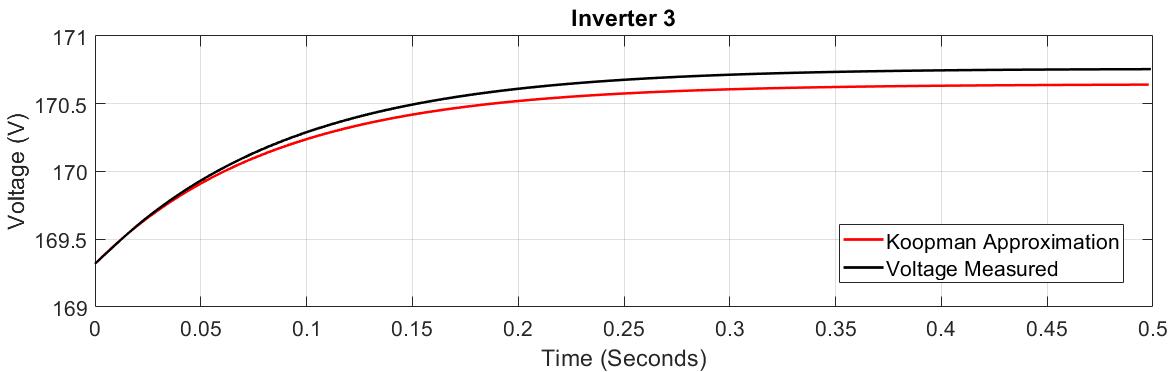}
	\includegraphics[width=1.0\linewidth]{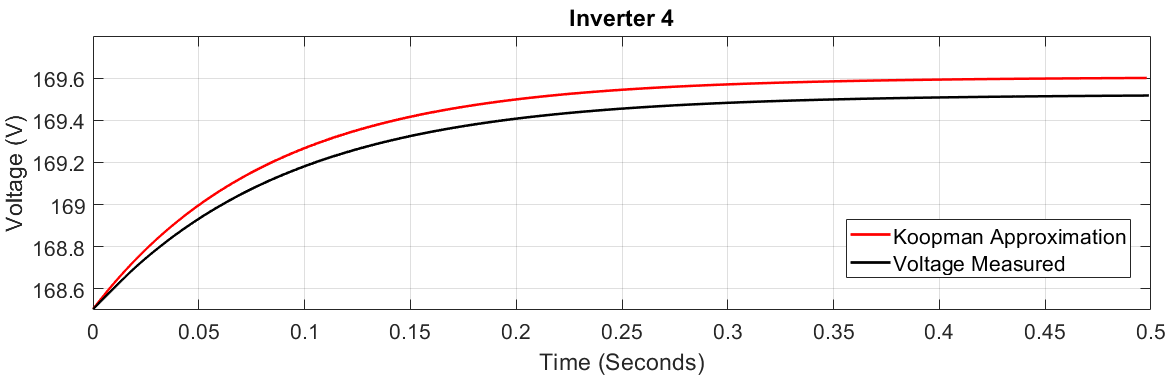}
	\includegraphics[width=1.0\linewidth]{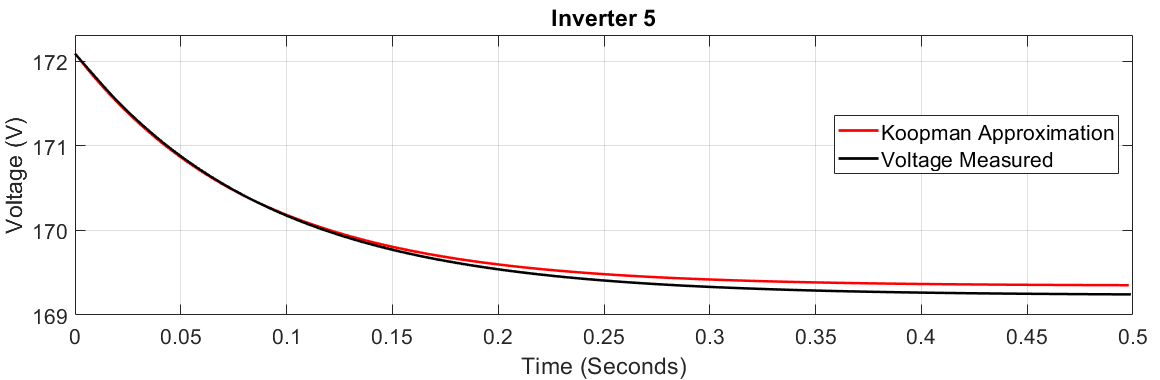}
	\caption{Koopman approximation for each inverter-based generator. Figures show the comparison between the voltage measured at the output of each inverter and the signal generated by the Koopman representation.}
	\label{Fig:Koopmaneachinverter}
\end{figure}

Thus, using the base defined by 
$$\mathbf{\Psi}=[V_{i} \quad V_{j} \quad V_{i}-V_{j} \quad (V_{i}-V_{j})^2]^{\top},$$ 
with the initial condition given by
$$\mathbf{\Psi}_{0}=[V_{i}(0) \quad V_{j}(0) \quad V_{i}(0)-V_{j}(0) \quad (V_{i}(0)-V_{j}(0))^2]^{\top},$$
and the matrices $\mathcal{A}_{i}$ and $\mathcal{B}_{i}$. The trajectories generated for each inverter are compared with the data measured as shown in Figure \ref{Fig:Koopmaneachinverter}. Notice that the approximation is good enough in the first steps, which makes it suitable for MPC.

\begin{figure}[t]
	\centering
	\includegraphics[width=1.0\linewidth]{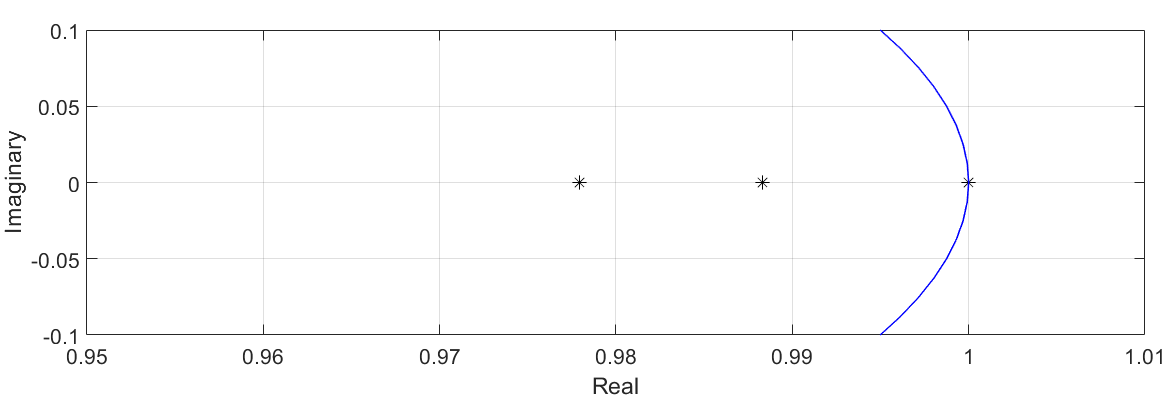}
	\caption{Eigenvalue plotting for all inverter's state matrices in the unit circle and zoom of the eigenvalues close to 1. There is a zero eigenvalue, two eigenvalues close to one, and one eigenvalue exactly in one.}
	\label{Fig:unitcircle}
\end{figure}

The discrete system described by $\mathcal{A}_{i}$ has $n$ discrete eigenvalues plotted in the unit circle as shown in Figure~\ref{Fig:unitcircle}; most of them are close to 1 and there is a zero eigenvalue. 

\begin{figure}[htp]
	\centering
	\includegraphics[width=1.0\linewidth]{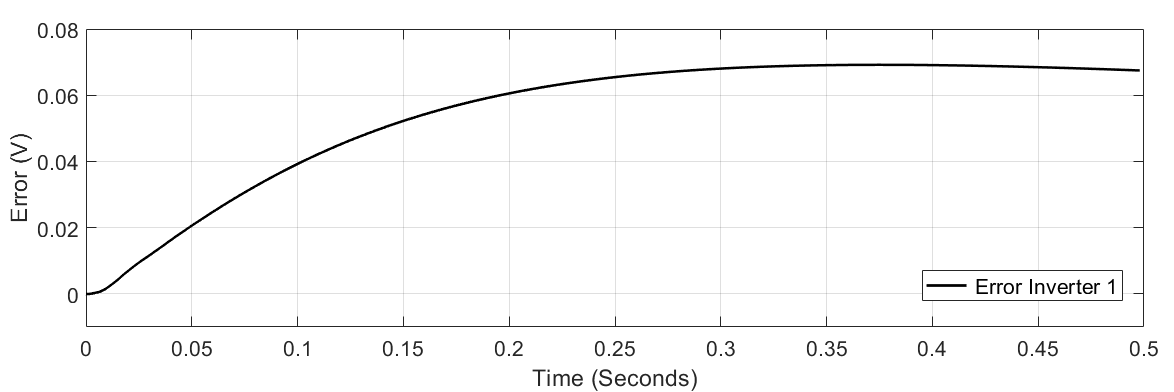}
	\caption{Error betweeen the data-measured and the Koopman approximation for inverter one.}
	\label{Fig:errorgenerator1}
\end{figure}

The comparison between the real measurements and the ones generated by using the Koopman approximation with matrices $\mathcal{A}$, $\mathcal{B}$, and $\mathcal{C}$ is made by plotting the error as shown in Figure~\ref{Fig:errorgenerator1}. The error is inferior to 1 percent in the first 0.5 seconds, which is suitable for the MPC design. 

\subsection{Load Changing Simulation}
Loads at nodes 3, 5, 6 , 9, 14 change from 0 to 1kW and 1kVar at $t=5$s.

\begin{figure}[htp]
	\centering
	\includegraphics[width=1.0\linewidth]{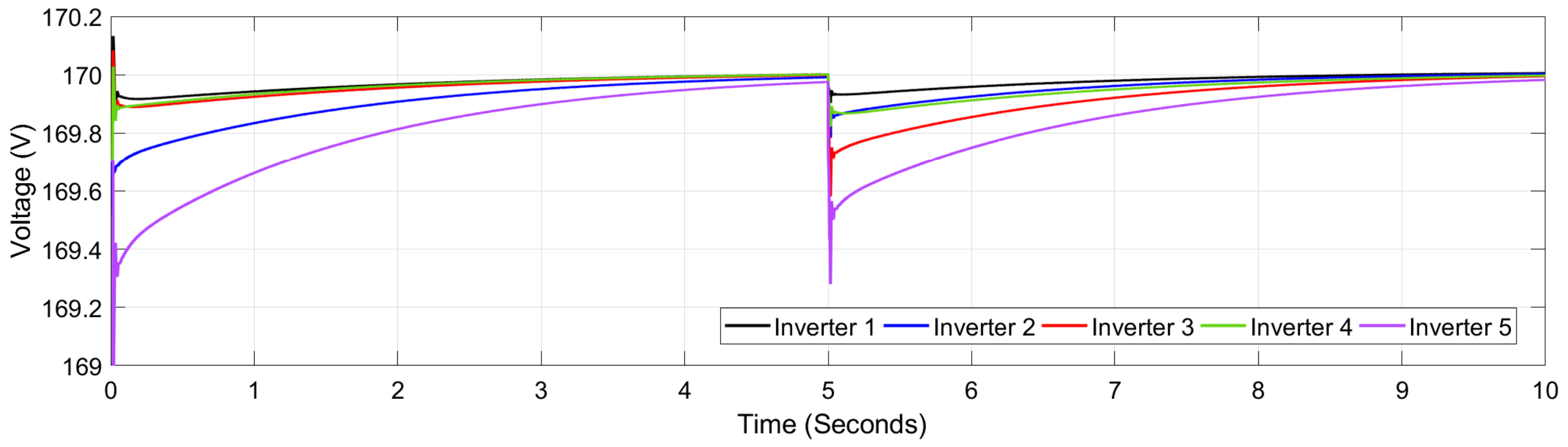}
	\caption{Voltage measured at the output of each inverter with load changes at $t=0$s and $t=5$s.}
	\label{Fig:loadchangeallvoltages}
\end{figure}

\begin{figure}[htp]
	\centering
	\includegraphics[width=1.0\linewidth]{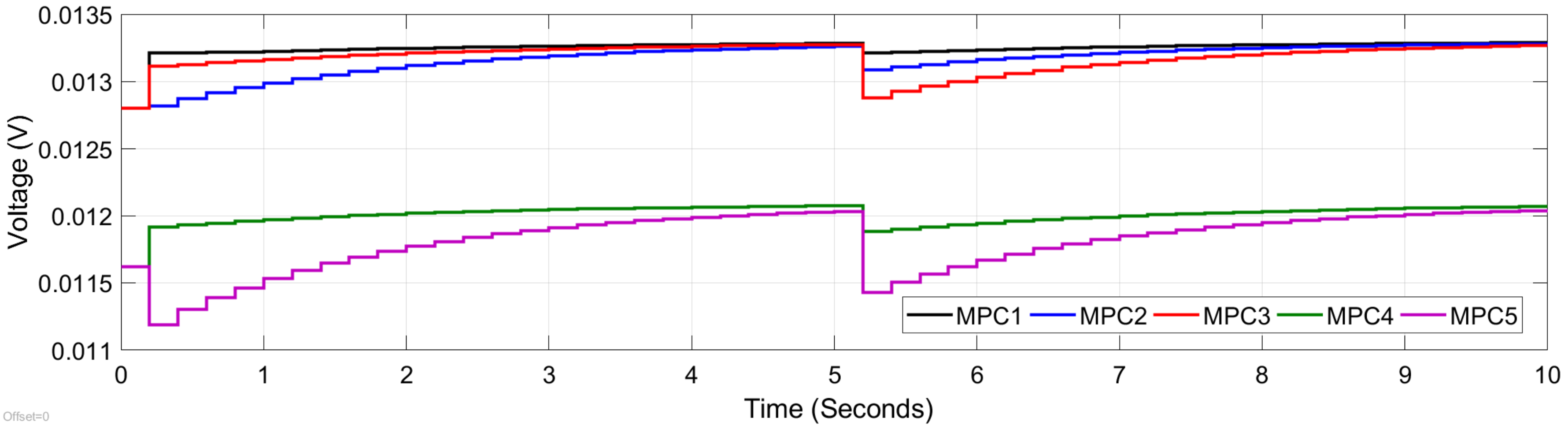}
	\caption{The MPC signals act at $t=0$s at the starting of the system, then at $t=5$s the signals changes until the voltage of reference is reached.}
	\label{Fig:loadchangealMPC}
\end{figure}

\begin{figure}[htp]
	\centering
	\includegraphics[width=1.0\linewidth]{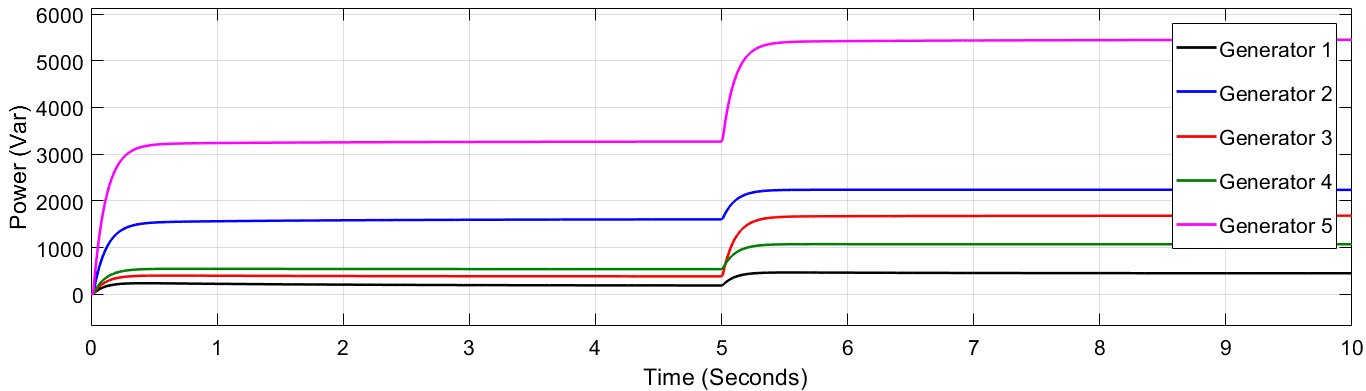}
	\caption{Reactive power measured at the output of each inverter when aditional load are connected at $t=5$s}
	\label{Fig:loadchangealReactivePowerAll}
\end{figure}

The voltage measured at the output of each inverter-based generator is shown in Figure~\ref{Fig:loadchangeallvoltages}. Voltage variations appear first at $t=0$s due to the initial conditions of the system, and at $t=5$s when loads at nodes 3, 5, 6, 9, and 14 are connected. The controller regulates the voltage after almost five seconds with some little oscillations as shown in Figure~\ref{Fig:loadchangealMPC}. The reactive power measured at each converter is shown in Figure~\ref{Fig:loadchangealReactivePowerAll}; the values change when loads are connected at $t=5$s. The reactive power supplied by each inverter not only depends on the maximum power rating but also on the transmission line's impedance values. The controller can regulate voltage while keeping the power-sharing condition.

\subsection{Transmission Line Changing Simulation}
\begin{figure}[htp]
	\centering
	\includegraphics[width=1.0\linewidth]{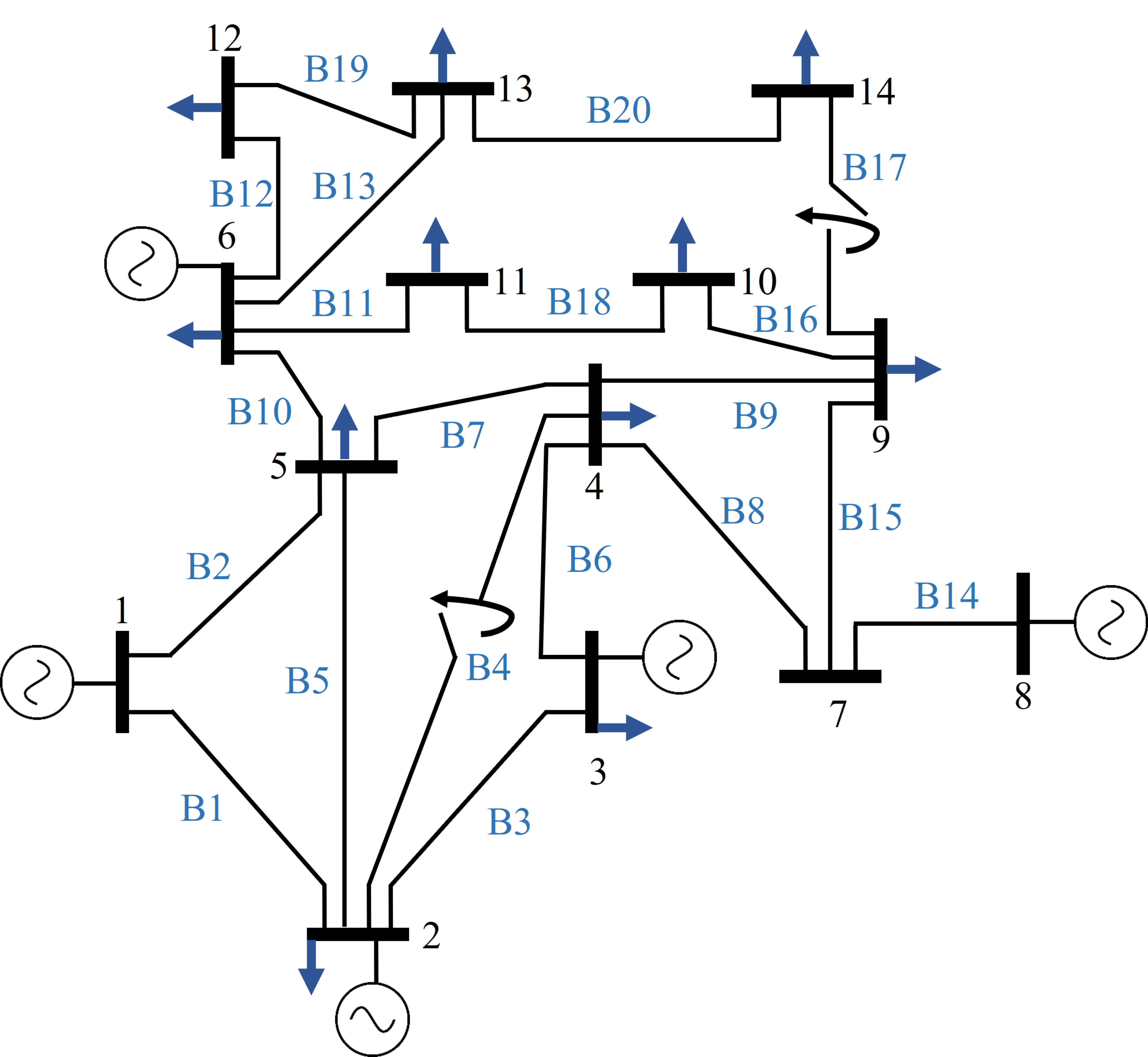}
	\caption{Transmission line changes between nodes two and four, and nodes nine and fourteen.}
	\label{Fig:transmisionlinefailure}
\end{figure}

Contingencies (disconnection of branches) between transmission lines are simulated by connecting nodes two and four, and nine and fourteen, as shown in Figure~\ref{Fig:transmisionlinefailure}. At $t=0$s nodes two and four and nine and fourteen are disconnected and the interrupters are opened; then at $t=5$s, the interrupters are closed, connecting nodes two and four and nine and fourteen through the transmission lines B4 and B17, respectively.

\begin{figure}[htp]
	\centering
	\includegraphics[width=1.0\linewidth]{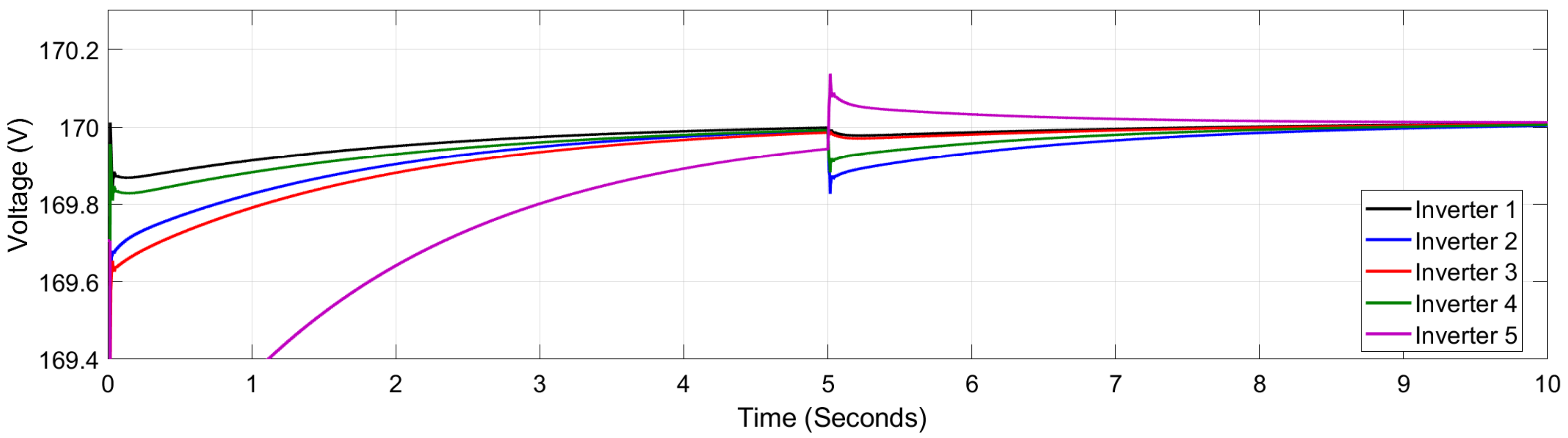}
	\caption{Zoom of the voltage measured at each inverter's output after the transmission line configuration changing. }
	\label{Fig:2linechangeallvoltages}
\end{figure}

\begin{figure}[htp]
	\centering
	\includegraphics[width=1.0\linewidth]{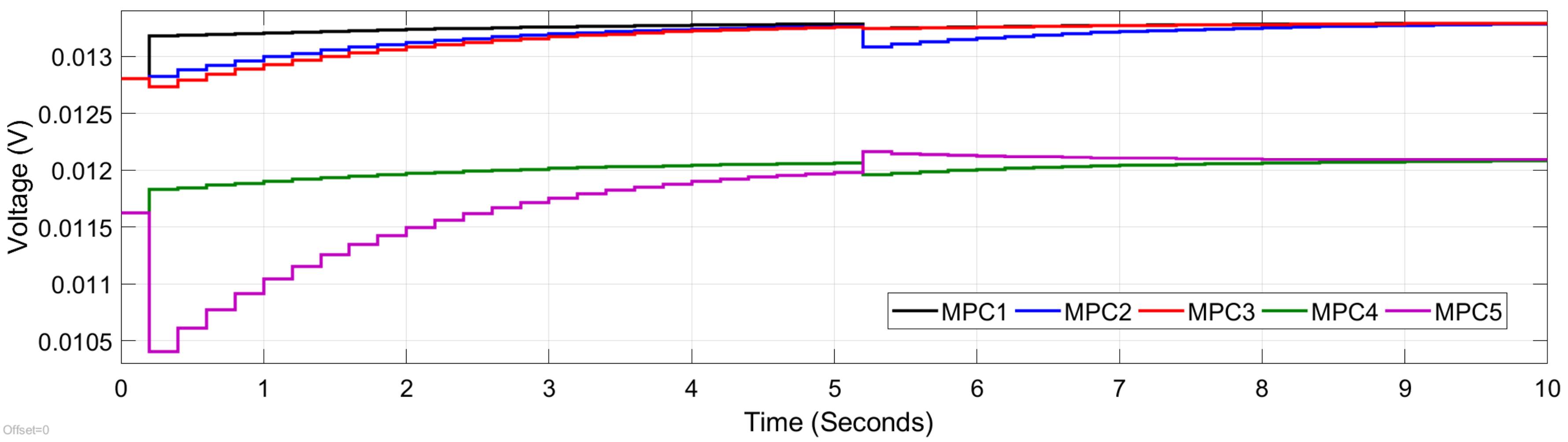}
	\caption{MPC signal for each inverter. After the connection at $t=0$s the signal is capable to regulate the voltage after transmission lines changes at $t=5$s.}
	\label{Fig:2linechangeMPC2}
\end{figure}

\begin{figure}[htp]
	\centering
	\includegraphics[width=1.0\linewidth]{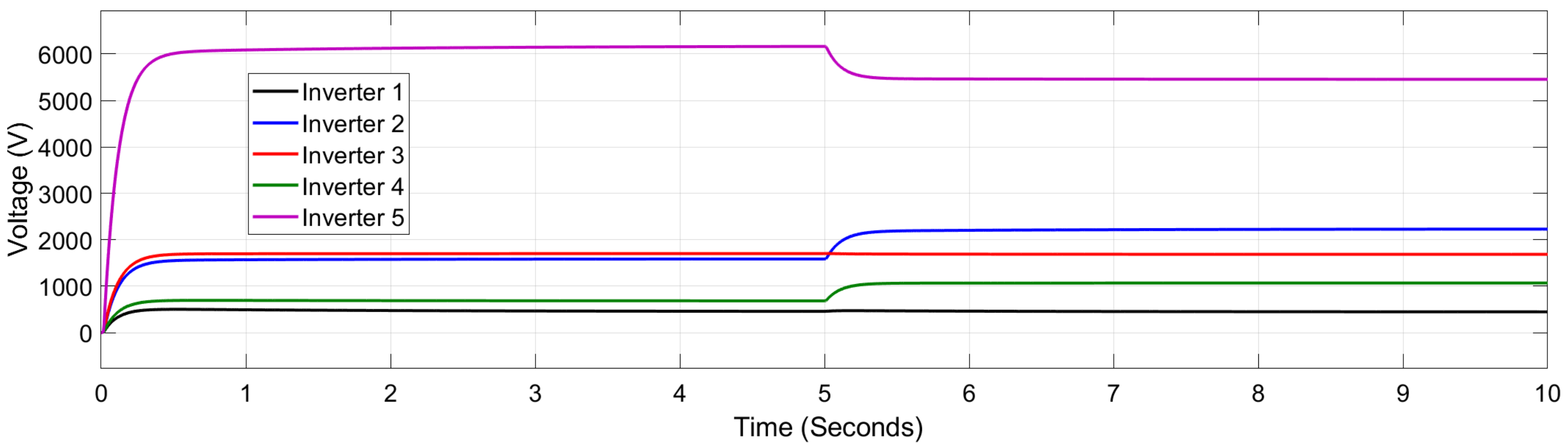}
	\caption{Reactive power supplied for each inverter. Notice the change in the reactive power after the changes in the transmission lines configuration at $t=5$s.}
	\label{Fig:2linechangeReactivePowerAll}
\end{figure}

The voltage measured at each inverter with zoom at $t=5$s is shown in Figure~\ref{Fig:2linechangeallvoltages}. Switches are closed in transmission lines B4 and B17, connecting inverters 2 and 4, and 9 and 14, respectively. Voltages stay around the reference value keeping the power-sharing condition. The controller takes less than 5 seconds to reach the desired condition as it is shown in Figure~\ref{Fig:2linechangeMPC2} where the MPC signals for each inverter appear. The reactive power supplied by each inverter, after the transmission line reconfiguration at $t=5$s, is shown in Figure~\ref{Fig:2linechangeReactivePowerAll} the power-sharing among sources also changes.

\subsection{Communication Graph Changing Simulation}
\begin{figure}[htp]
	\centering
	\includegraphics[width=1.0\linewidth]{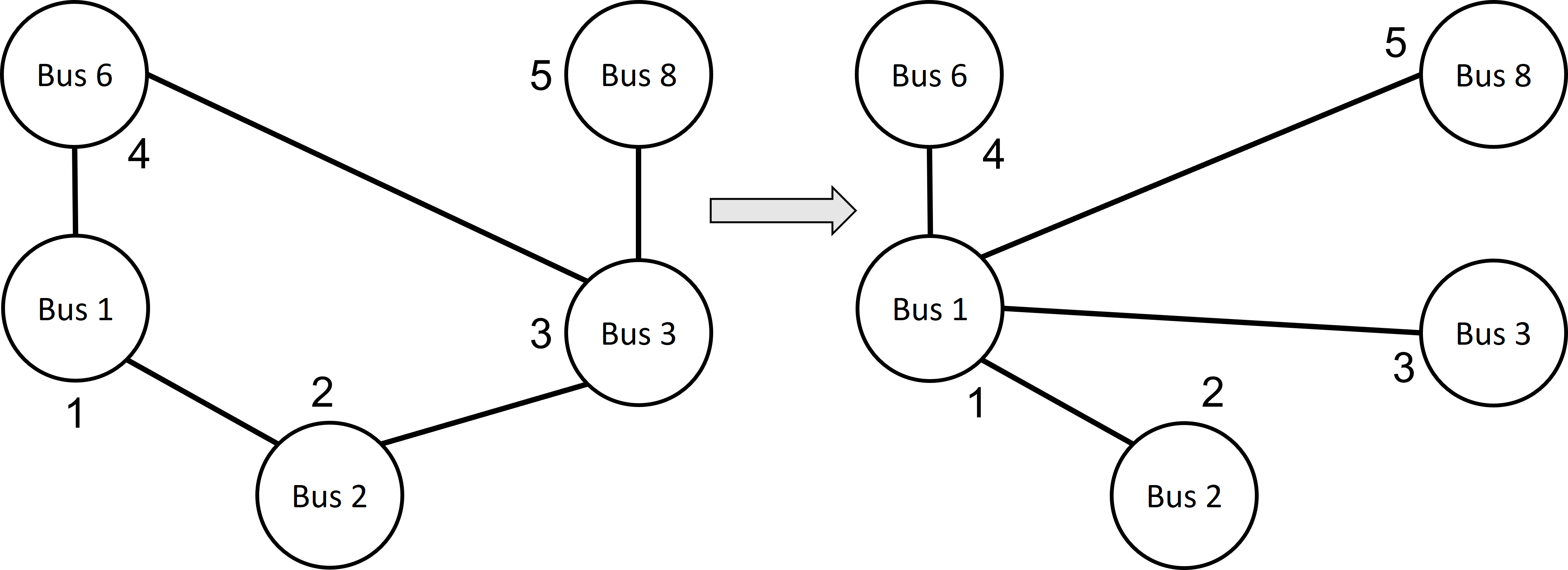}
	\caption{Communication graph changing}
	\label{Fig:GraphChanging}
\end{figure}

For this scenario, the communications graph among inverters changes at $t=5$s, as it is shown in Figure~\ref{Fig:GraphChanging}. For both cases, the condition of having at least a spanning tree for consensus reaching is guaranteed.

\begin{equation*}
	\mathcal{L}_{2}=\left[ 
	\begin{array}{ccccc}
		4  & -1  & -1 & -1 & -1 \\
		-1 &  1  & 0  &  0 & 0 \\
		-1 &  0  &  1 &  0 & 0  \\
		-1 &  0  & 0  & 1  & 0 \\
		-1 &  0  & 0  & 0 &  1
	\end{array}
	\right]
\end{equation*}

\begin{figure}[htp]
	\centering
	\includegraphics[width=1.0\linewidth]{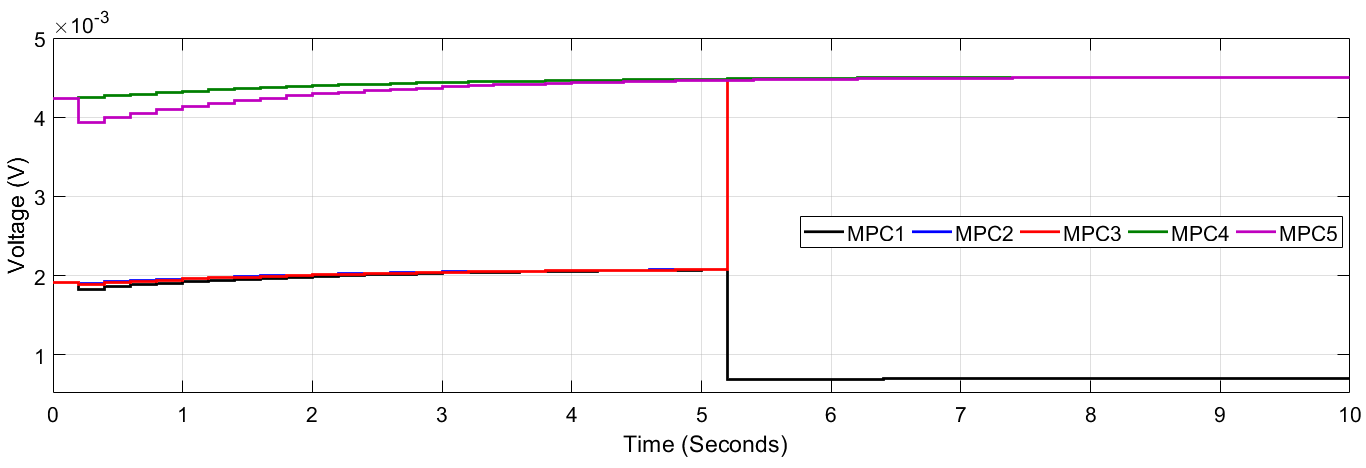}
	\caption{MPC signals at each inverter. The loads are connected at $t=0$s and the graph communication changes at $t=5$s.}
	\label{Fig:3graphchangeMPC}
\end{figure}

\begin{figure}[htp]
	\centering
	\includegraphics[width=1.0\linewidth]{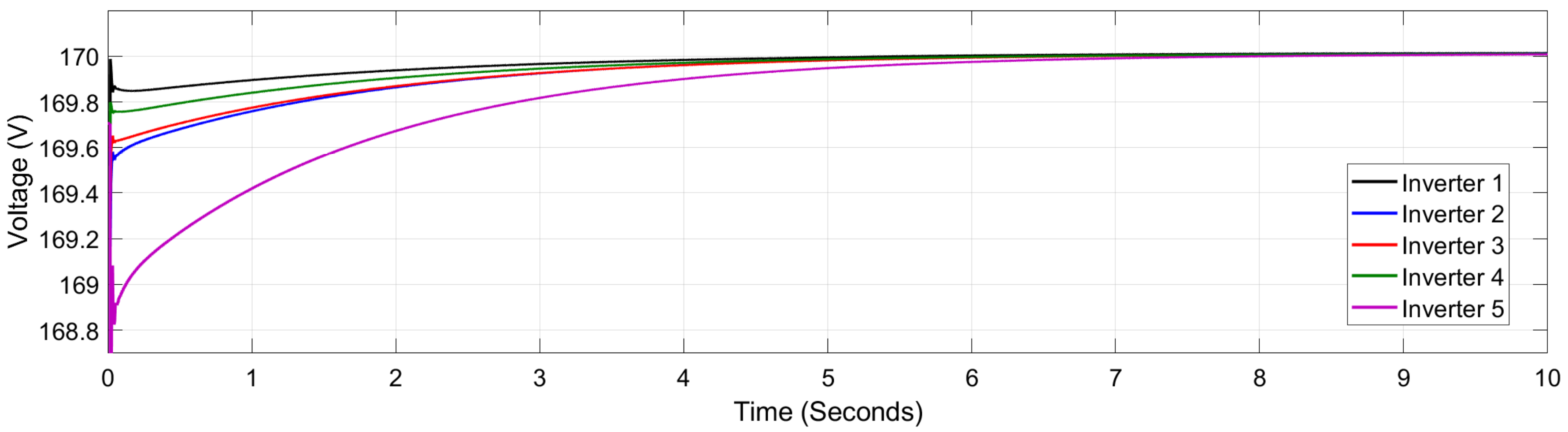}
	\caption{Voltage measured at each inverter output, the voltage variation at $t=5$s is not notorious after the graph changing.}
	\label{Fig:3graphchangevoltages}
\end{figure}

\begin{figure}[htp]
	\centering
	\includegraphics[width=1.0\linewidth]{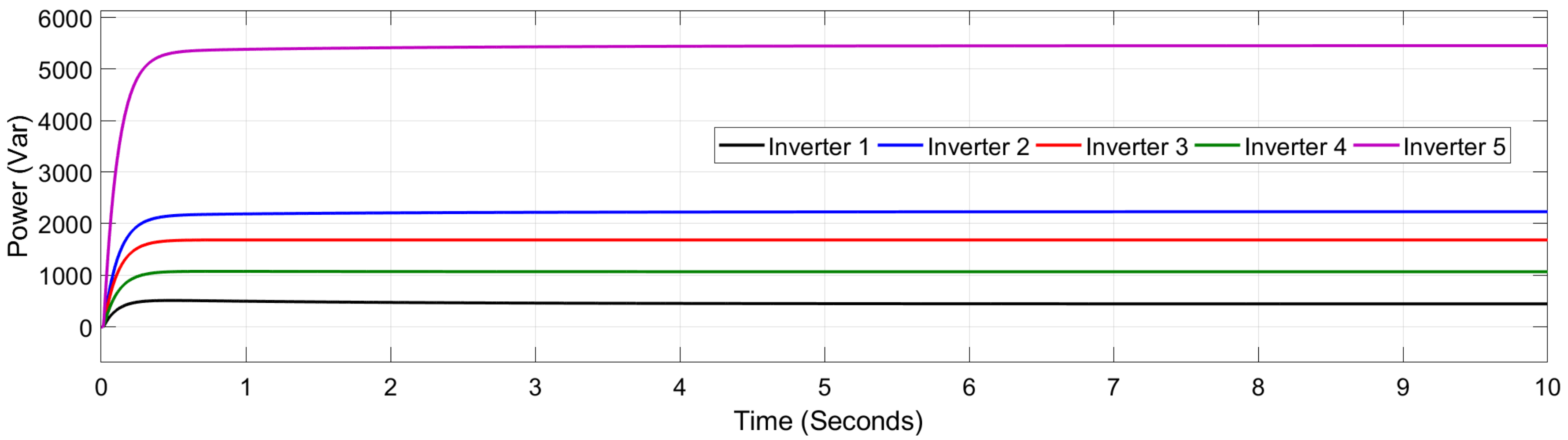}
	\caption{Reactive power measured at each inverter when the graph changes at $t=5$s.}
	\label{Fig:3graphchangeReactivePower}
\end{figure}

The predictive control signal is shown in Figure~\ref{Fig:3graphchangeMPC} where the signal changes at $t=5$s,  the magnitude of the signal shows some changes in inverter 1 and inverter 3. The output voltage at each inverter is presented in Figure.~\ref{Fig:3graphchangevoltages}, voltage variations are not significative when the communication graph changes at $t=5$s, each inverter reaches the voltage of reference. Finally, the reactive power at each inverter is shown in Figure~\ref{Fig:3graphchangeReactivePower} where there is not variation at $t=5$s when the graph changes.

\subsection{Algorithm Comparison with Nonlinear MPC}
\begin{figure}[htp]
	\centering
	\includegraphics[width=1.0\linewidth]{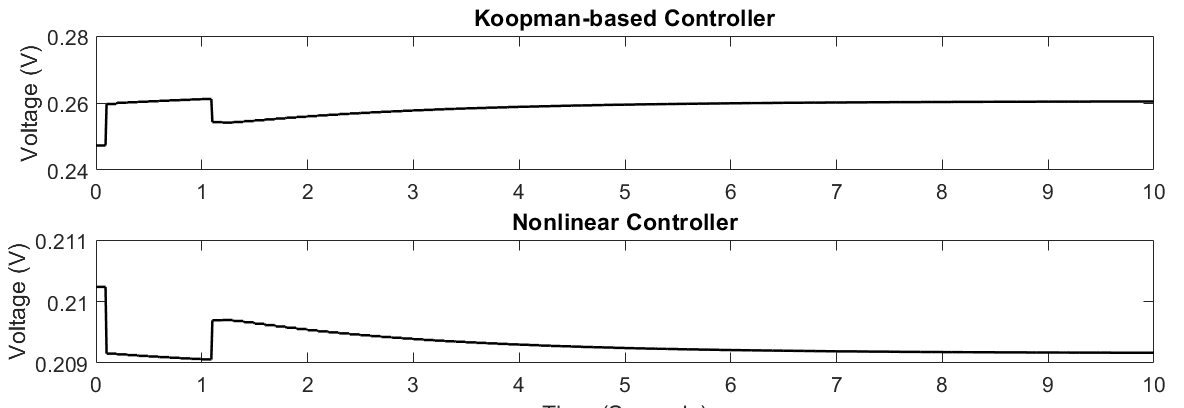}
	\caption{Comparison between the MPC signals generated by the Koopman -based and the nonlinear-based approaches.}
	\label{Fig:mpcComparisonNL}
\end{figure}

\begin{figure}[htp]
	\centering
	\includegraphics[width=1.0\linewidth]{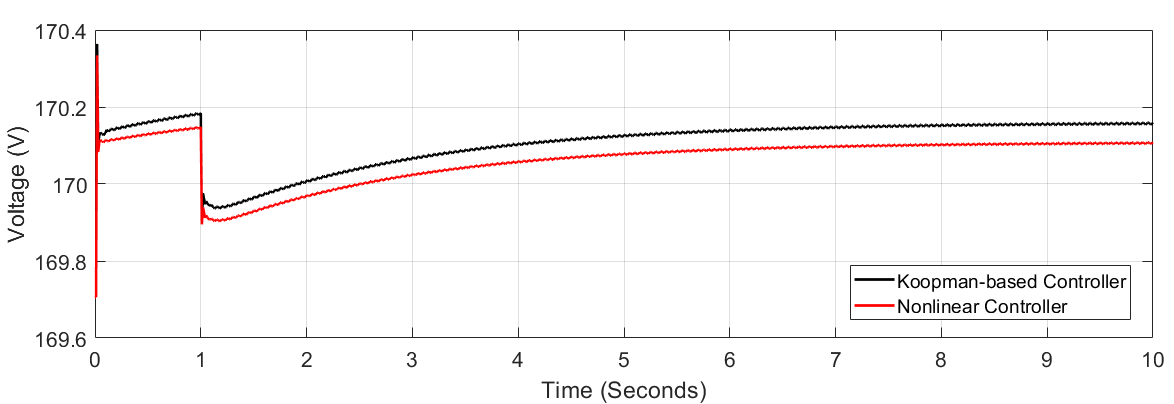}
	\caption{Output voltage at inverter one generated by using the Koopman-based and the nonlinear-based MPC.}
	\label{Fig:comparisonKNL}
\end{figure}

The control signals generated under each approach are shown in Figure~\ref{Fig:mpcComparisonNL}. Each signal reaches a different value at steady state which can be seen in the difference between the output voltages shown in Figure~\ref{Fig:comparisonKNL}.

\begin{table}[htp]
\caption{Comparison Beetween Koopman-based and Nonlinear MPC algorithms}
	\begin{tabular}{lcc}
		\hline
		& \begin{tabular}[c]{@{}c@{}}Koopman-Based \\ Algorithm\end{tabular} & \begin{tabular}[c]{@{}c@{}}Non-linear \\ Algorithm\end{tabular} \\ \hline
		Sampling Time       & 0.01s                                                              & 0.01s                                                           \\ 
		Prediction Horizon  & 10                                                                 & 10                                                              \\ 
		Q                   & 1                                                                & 1                                                             \\ 
		R                   & 1                                                                  & 1                                                               \\ 
		Solver              & Fmincon                                                             & Fmincon                                                         \\ 
		$V_{i}$, $V_{j}$     & $ [170,171]$                                                             & $[170,171]$                                                         \\ 
		Time used per cycle & 0.2098                                                             & 0.5700                                                          \\ \hline
	\end{tabular}
\label{Table:algorithmcompa}
\end{table}

A comparison between the nonlinear and the Koopman approaches is presented in Table~\ref{Table:algorithmcompa}. Both algorithms use identical parameters and the same nonlinear solver to compare the time used per cycle for solving the optimization problem. It is clear that the Koopman-based algorithm uses less time for solving the problem, being about 271,69 percent faster than the one using the nonlinear algorithm. 

\begin{figure}[htp]
	\centering
	\includegraphics[width=1.0\linewidth]{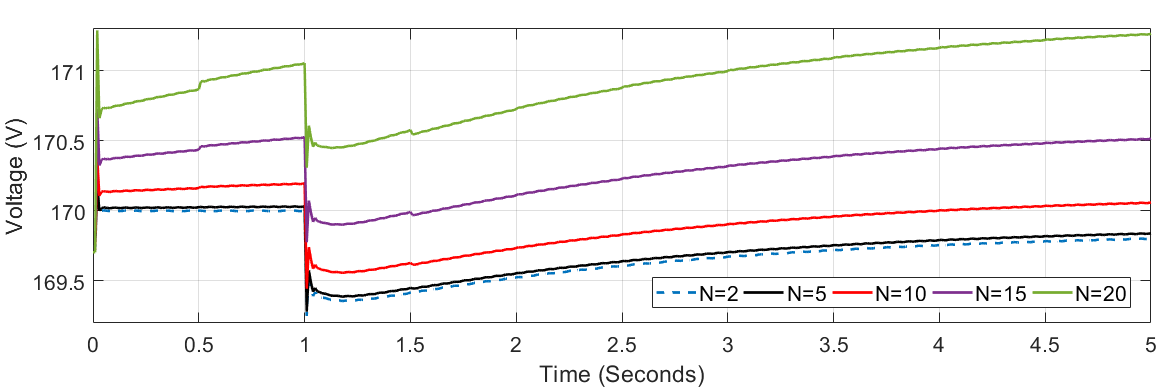}
	\caption{Comparison of the voltage output at inverter one for different prediction horizons.}
	\label{Fig:differenthorizon}
\end{figure}

Finally, a comparison of the voltage measured at inverter one by changing the prediction horizon $H_{p}=[2 \quad 5 \quad 10 \quad 15 \quad 20]$ is shown in Fig~\ref{Fig:differenthorizon}, the prediction horizon that best approximate to the reference value is $H_{p}=10$. 

\section{Conclusion and Future Work}
A distributed MPC using a Koopman-based approach was presented. The dynamic of the inverters with droop-control were represented in the Koopman lifted space by using the EDMD algorithm. Thus, the linear representation in the Koopman-space of the inverters was used instead of the nonlinear one with a quadratic voltage term. The distributed problem was defined for the restrictions of the optimization problem by including a term with the Laplacian matrix. The convergence of the algorithm was guaranteed by the selection of the weight matrices. The simulation shows that the MPC, using the Koopman representation, regulates the voltage of each inverter in the MG when load changes, transmission line changes, and communication graph changes occur. For future work, the same problem can be set for frequency control without assuming a decoupling between voltage and frequency.

\section{Appendix}
In this section, the lifted matrices $\mathcal{A}_{1}$, $\mathcal{B}_{1}$, and $\mathcal{C}_{1}$ are presented.

\begin{equation*}
	\mathcal{A}_{1}=\left[ 
	\begin{array}{cccc}
		0.6628 &   0.3372 &   0.3256 &   0.0001 \\
		0.3334 &   0.6666 &  -0.3332 &  -0.0000 \\
		0.3294 &  -0.3294 &   0.6588 &   0.0002 \\
		0.0017 &  -0.0016 &   0.0033 &   0.9780
	\end{array}
	\right];
	\qquad
\end{equation*}
\begin{equation*}
	\mathcal{B}_{1}=\left[ 
	\begin{array}{c}
		0.0084\\
	   -0.0002\\
		0.0085\\
		-0.0067
	\end{array}
	\right];
\end{equation*}

\begin{equation*}
	\mathcal{C}_{1}=\left[
	\begin{array}{cccc}
		0.6667  &  0.3333  &  0.3333  &  0.0000
	\end{array}
	\right].
\end{equation*}

\bibliographystyle{elsarticle-num} 
\bibliography{mgkoopman}
\end{document}